\documentclass[aps,prl,superscriptaddress,longbibliography,%pdflatex,
floatfix,tightenlines,twocolumn]{revtex4-2}%,twocolumn]{revtex4}
\usepackage{subfigure}
\usepackage{amsmath}
\usepackage{amssymb}
\usepackage{amsfonts}
\usepackage{epsfig}
\usepackage[T1]{fontenc}
\usepackage[dvips]{color}
\usepackage[colorlinks,bookmarks=false,citecolor=magenta,linkcolor=magenta,urlcolor=magenta]{hyperref}
\usepackage{bm}% bold math
\usepackage{times}
\usepackage{amsthm}
\definecolor{orange}{cmyk}{0,0.3,1,0.2}

\renewcommand{\figurename}{Fig.}
\begin{document}
\title{
Metrological characterisation of  non-Gaussian  entangled states of superconducting qubits
}
\author{Kai Xu}
\thanks{These authors contributed equally to this work.}
\affiliation{Institute of Physics, Chinese Academy of Sciences, Beijing 100190, China}
\author{Yu-Ran Zhang}
\thanks{These authors contributed equally to this work.}
	\affiliation{Theoretical Quantum Physics Laboratory, RIKEN Cluster for Pioneering Research, Wako-shi, Saitama 351-0198, Japan}
  \author{Zheng-Hang Sun}
  \thanks{These authors contributed equally to this work.}
  \affiliation{Institute of Physics, Chinese Academy of Sciences, Beijing 100190, China}

  \author{Hekang Li}
  \affiliation{Institute of Physics, Chinese Academy of Sciences, Beijing 100190, China}

  \author{Pengtao Song}
  \affiliation{Institute of Physics, Chinese Academy of Sciences, Beijing 100190, China}

    \author{Zhongcheng Xiang}
\affiliation{Institute of Physics, Chinese Academy of Sciences, Beijing 100190, China}

  \author{Kaixuan Huang}
  \affiliation{Institute of Physics, Chinese Academy of Sciences, Beijing 100190, China}

  \author{Hao Li}
  \affiliation{Institute of Physics, Chinese Academy of Sciences, Beijing 100190, China}

  \author{Yun-Hao Shi}
  \affiliation{Institute of Physics, Chinese Academy of Sciences, Beijing 100190, China}

  \author{Chi-Tong Chen}
  \affiliation{Institute of Physics, Chinese Academy of Sciences, Beijing 100190, China}
  \author{Xiaohui Song}
    %\email{dzheng@iphy.ac.cn}
  \affiliation{Institute of Physics, Chinese Academy of Sciences, Beijing 100190, China}
  \author{Dongning Zheng}
    %\email{dzheng@iphy.ac.cn}
  \affiliation{Institute of Physics, Chinese Academy of Sciences, Beijing 100190, China}
  %\affiliation{CAS Centre for Excellence in Topological Quantum Computation, University of Chinese Academy of Sciences, Beijing 100190, China}
	\author{Franco Nori}
	\email{fnori@riken.jp}
		\affiliation{Theoretical Quantum Physics Laboratory, RIKEN Cluster for Pioneering Research, Wako-shi, Saitama 351-0198, Japan}
	  	\affiliation{Physics Department, University of Michigan, Ann Arbor, Michigan 48109-1040, USA}
  \author{H. Wang}
  \email{hhwang@zju.edu.cn}
  \affiliation{Interdisciplinary Centre for Quantum Information, State
Key Laboratory of Modern Optical Instrumentation, and
Zhejiang Province Key Laboratory of Quantum Technology
and Device, Department of Physics, Zhejiang University,
Hangzhou 310027, China}
\author{Heng Fan}
\email{hfan@iphy.ac.cn}
\affiliation{Institute of Physics, Chinese Academy of Sciences, Beijing 100190, China}
\affiliation{CAS Centre for Excellence in Topological Quantum Computation, UCAS, Beijing 100190, China}


\begin{abstract}
Multipartite entangled states are significant resources for both quantum information processing and quantum metrology
\cite{Giovannetti2011}.
In particular, non-Gaussian entangled states are predicted to achieve a higher sensitivity of precision measurements than
 Gaussian states \cite{Pezze2018}.
%but not all the entangled states
%, algebraic separability properties being cast aside,
%are equally useful for developing quantum protocols that outperforms classical ones \cite{Pezze2018}.
On the basis of metrological sensitivity, the conventional linear Ramsey squeezing parameter (RSP) \cite{Wineland1994,Ma2011}
efficiently characterises the Gaussian entangled atomic states but fails for much wider classes of highly sensitive
non-Gaussian states. These complex non-Gaussian entangled states can be classified by the nonlinear squeezing
parameter (NLSP), as a generalisation of the RSP with respect to nonlinear observables \cite{Gessner2019},
and identified via the Fisher information \cite{BRAUNSTEIN1994}. However, the NLSP has never been measured
experimentally.
%, and the extraction of the Fisher information
%via performing a full phase estimation experiment
%indicates the sensitivity of precision measurement in a system.
Using a 19-qubit programmable superconducting processor, here we report the characterisation of multiparticle %Gaussian and non-Gaussian
entangled states generated during its nonlinear dynamics. First, selecting 10 qubits, we measure the RSP and
the NLSP by single-shot readouts of collective spin operators in several different directions. Then, by extracting
the Fisher information of the time-evolved state of all 19 qubits, we observe a large metrological gain
of $9.89^{+0.28}_{-0.29}$~dB over the standard quantum limit \cite{Giovannetti2011}, indicating a
high level of multiparticle entanglement for quantum-enhanced phase sensitivity.
%Merging different concepts of entanglement witnesses, our experiments
%classify the complicated multiparticle entangled states without quantum state tomography.
Benefiting
from high-fidelity full controls and addressable single-shot readouts, the superconducting processor with
interconnected qubits provides an ideal platform for engineering and benchmarking non-Gaussian entangled states
that are useful for quantum-enhanced metrology.
%and realising other
%complex spin-squeezing algorithms \cite{KITAGAWA1993}.
\end{abstract}

\maketitle

The ability to create and manipulate the entangled states of multiparticle quantum systems is crucial
for advanced quantum technologies, including quantum metrology \cite{Giovannetti2011}, quantum error correction
\cite{%Knill2005,
Campbell2017}, quantum communications \cite{Zhao2004,Wehner2018}, quantum simulations \cite{Georgescu2014},
and fundamental tests of quantum theory \cite{Wang2020}.
A universal quantum   computer %\cite{Deutsch1985}
is able to deterministically generate multiparticle entangled
states with numerous sequences of single- and two-qubit operations. However, the conventional step-by-step method is
very challenging to scale up and increases exposure to noise.
Instead,  parallel entangling operations, involving all-to-all connectivity, can efficiently create various types
of entangled states, and have also been suggested to obtain polynomial or exponential speedups in
some quantum algorithms and quantum simulation \cite{Lu2019,Figgatt2019}. Realised via the free evolution under a one-axis twisting
(OAT) Hamiltonian, the parallel entangling operation first transforms the initial coherent spin state to
squeezed spin states \cite{Ma2011,KITAGAWA1993} and then to non-Gaussian entangled states \cite{Pezze2018}, including multicomponent
atomic Schr\"{o}dinger cat states  and the GHZ state \cite{Song2019}. In the squeezed regime, squeezing of a collective spin, described
by Gaussian statistics, represents the improvement of phase sensitivity to SU(2) rotations over the standard quantum limit
\cite{Giovannetti2011,Ma2011} and can be characterised by the Ramsey squeezing parameter (RSP) $\xi^2_{\textrm{R}}$ \cite{Wineland1994}.
 In the over-squeezed regime, multipartite entanglement of the non-Gaussian spin states can be witnessed by  extracting the
 Fisher information  $F$ \cite{BRAUNSTEIN1994}, related to the phase sensitivity in Ramsey interferometry via the Cram\'{e}r-Rao
 bound $(\Delta\theta)^2\geq1/F$ \cite{Giovannetti2011,Ma2011}.
 %\cite{Cramer1946}.
 Furthermore, the non-Gaussian entangled states can be classified  by the nonlinear
 squeezing parameter (NLSP) $\xi^2_{\textrm{NL}}$ \cite{Gessner2019}, extending the concept of spin squeezing to nonlinear observables.
Despite many achievements  in generating linear spin squeezing (e.g., in Bose-Einstein condensates %(BECs)
\cite{Orzel2001,Esteve:2008aa,Gross:2010aa,Riedel:2010aa,Luo2017,Zou2018},
atomic ensembles \cite{Fernholz2008,Chaudhury2007,Appel2009,Takano2010,Leroux2010,
Schleier-Smith2010,Hamley2012,Sewell2012,Bohnet2014,McConnell2015,Hosten2016,Hosten2016a},
and trapped ions \cite{Bohnet2016}), the %generation and the identification of
non-Gaussian entangled states, believed to perform higher-sensitive quantum phase estimation, quantum simulations
\cite{Georgescu2014} and classically intractable quantum algorithms, %(e.g., Shor's algorithm \cite{Shor1997}),
are
attracting growing interests \cite{Parigi2007,Haas2014,Strobel2014,Andersen2015}.

 Here, we measure the  NLSP using 10 interconnected superconducting qubits, which
 requires the capability of single-shot readouts of collective spin operators. Compared with the linear RSP  and the
 Fisher information, our experiments help to analyse different classes of complicated non-Gaussian entangled states
 during the OAT evolution of the multi-qubit state. Moreover, by extracting the Fisher information, our experiments
 achieve a metrological gain, $F/N=9.89^{+0.28}_{-0.29}$~dB using $N=19$ qubits, which is larger than those
 obtained in many other experimental platforms with a much larger number of  particles. Our work will stimulate
 interests in non-Gaussian entangled states of quantum many-body systems, with further applications in
 practical quantum metrology and quantum information processing.

 \begin{figure}[t]
\centering
\includegraphics[width=0.99\linewidth]{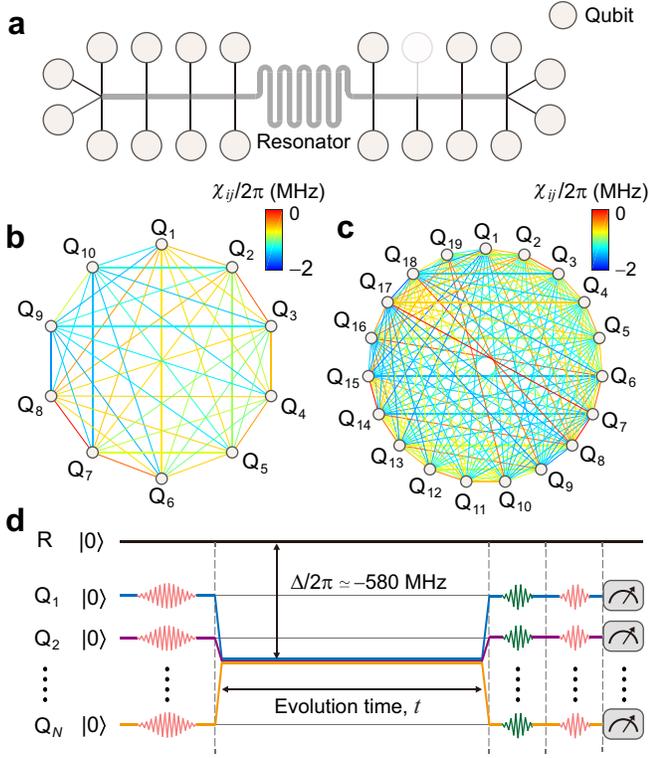}
\caption{\textbf{Superconducting quantum processor and experimental pulse sequence.} \textbf{a}, Simplified schematic of the
superconducting quantum processor, showing 19 qubits interconnected by the central bus resonator R.
\textbf{b}, \textbf{c}, Effective all-to-all coupling strengths, $\chi_{ij}$, for (\textbf{b}) selected 10 qubits and (\textbf{c}) all 19 qubits.
\textbf{d}, Experimental waveform pulse sequence. Ten or nineteen qubits are initially prepared at $|0\rangle$ at their idle points,
and then transformed to $|+\rangle$ by a collective Y$_{\frac{\pi}{2}}$ gate. After the free evolution with a time $t$, when all qubits
are equally detuned to the resonator R with $\Delta/2\pi\simeq-580$~MHz, all qubits at their idle points are measured in the
same direction.}\label{fig1}
\end{figure}

 In our experiments, 19 addressable transmon qubits (Q$_j$, with $j$ varied from 1 to 19), capacitively coupled to a
 resonator bus {R}, are chosen to effectively engineer a OAT Hamiltonian, see Fig.~\ref{fig1}a and
refs.~\cite{Song2019,Xu2020}. By equally detuning selected qubits from the resonator by
 $\Delta/2\pi\simeq-580$~MHz, the
 effective system Hamiltonian reads (we set $\hbar=1$)%, where $\hbar$ is the Planck's constant $h$ divided by $2\pi$)
 \begin{align}
 	\hat{H}=\sum_{1\leq i<j\leq N}\chi_{ij}(\hat{\sigma}_i^+\hat{\sigma}_j^-+h.c.),\label{ham}
 \end{align}
 where $\hat{\sigma}_j^+$  ($\hat{\sigma}_j^-$)  is the raising (lowering) operator of Q$_j$, and $\chi_{ij}$
 denotes the qubit-qubit coupling. As shown in Fig.~\ref{fig1}b,c for choosing 10 qubits (Q$_j$ with $j=1,2,\cdots,10$)
 and all 19 qubits, respectively, the effect of unbalanced qubit-qubit couplings, caused by the few cross-talk
 couplings between neighbouring qubits, can be ignored, see also refs.~\cite{Song2019,Xu2020}.
With $N$ selected  qubits	initialised at their idle points as $|00\cdots0\rangle_N$, we prepare these qubits in the
state $|++\cdots+\rangle_N$ via a Y$_{\frac{\pi}{2}}$ pulse, and then detune them equally from the resonator R
for the quench dynamics with a time $t$ before the readouts in the same direction (see the experimental pulse
sequence in Fig.~\ref{fig1}d). Our $N$-qubit system can be described by a family of linear collective spin operators
$\hat{\mathbf{J}}\equiv(\hat{J}_x,\hat{J}_y,\hat{J}_z)$, with
$\hat{J}_\beta\equiv\sum_{j=1}^N\hat{\sigma}_j^\beta/2$, and $\hat{\sigma}_j^\beta$ being Pauli matrices for
$\beta=x,y,z$. The Hamiltonian in Eq.~(\ref{ham}) can be approximately expressed as a OAT one
$\hat{H}\simeq-\chi \hat{J}_z^2$.

\begin{figure}[t]
\centering
\includegraphics[width=0.99\linewidth]{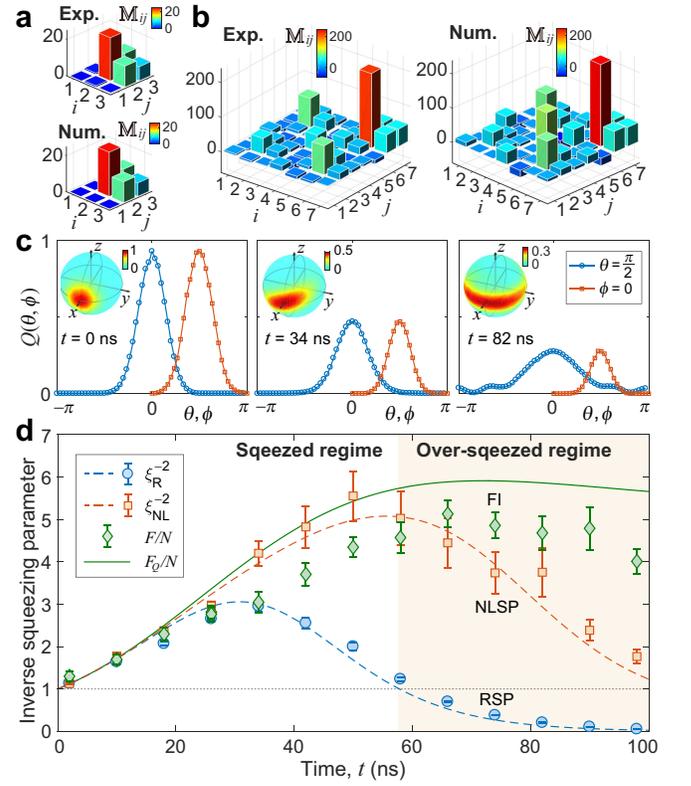}
\caption{\textbf{Linear Ramsey squeezing parameter versus nonlinear squeezing parameter for 10 superconducting
qubits.}
\textbf{a}, \textbf{b}, The measured matrices $\mathbb{M}$ to optimise (\textbf{a}) the linear RSP at $t=34$~ns, and (\textbf{b})
the NLSP at $t=50$~ns, respectively, compared with the numerical simulations.
\textbf{c}, Experimental data of the Husimi $Q$ functions of the states, $Q(\theta,\phi)$, against $\theta$ and $\phi$
at specific times with the rotations along the $x$-axis to widen the equatorial distributions.
At $t=0$~ns, 34~ns, and 82~ns, the $Q(\theta,\phi)$ represent the spin-coherent state, the spin-squeezed state,
and the non-Gaussian state, respectively. Insets: Experimentally measured $Q(\theta,\phi)$ of the evolved states
displayed in  spherical polar.
\textbf{d}, Time evolutions of the inverse linear RSP, $\xi^{-2}_{\textrm{R}}$, the inverse NLSP,
$\xi^{-2}_{\textrm{NL}}$, with a family of operators $\hat{\mathbf{S}}_{\textrm{exp}}$, and the normalised
Fisher information (FI), $F/N$,  compared with the numerical simulations without decoherence (dashed curves).
The green solid curve shows the numerical simulation of the normalised quantum Fisher information,
$F_Q/N=4\max_{\bm{n}\in\mathbb{R}^3}(\Delta_{\rho_t}\hat{J}_{\bm{n}})^2/N$, without decoherence,
which is the largest normalised Fisher information over all possible measurements and linear generators.
The error bars, indicating the standard deviations of the results, are calculated from 200,000 repetitive
experimental runs in total (see Supplementary Information for details).
}\label{fig2}
\end{figure}

We need to estimate an unknown parameter $\theta$, imprinted on the
time evolved state $\rho_t$  at time $t$ via unitary evolutions,
$\rho_t(\theta)=e^{-i\hat{J}_{\hat{n}}\theta}\rho_t e^{i\hat{J}_{\hat{n}}\theta}$,
with $\hat{J}_{\hat{n}}\equiv\hat{n}\cdot\hat{\mathbf{J}}$ being a collective spin operator in the direction
$\hat{n}\in \mathbb{R}^3$. For a family of $D$ accessible operators
$\hat{\mathbf{S}}=(\hat{S}_1,\hat{S}_2,\cdots,\hat{S}_D)$, the parameter $\theta$ is estimated from  the measurement
of the observable $\hat{S}_{\hat{m}}=\hat{m}\cdot\hat{\mathbf{S}}$, with $\hat{m}\in\mathbb{R}^D$,
 as a linear combination of  accessible operators. Then, the optimal metrological squeezing parameter of $\rho_t$ for
$\hat{\mathbf{S}}$ can be written as \cite{Gessner2019}
	 \begin{align}
	 	\xi^2_{\textrm{opt}}[\rho_t, \hat{\mathbf{S}}]=\min_{\hat{m}\in \mathbb{R}^D}\min_{\hat{n}\in\mathbb{R}^3}\frac{N(\Delta_{\rho_t}\hat{S}_{\hat{m}})^2}{|\langle[\hat{S}_{\hat{m}},\hat{J}_{\hat{n}}]\rangle_{\rho_t}|^2},
	 \end{align}
where ${(\Delta_{\rho}\hat{O}})^2\equiv \langle\hat{O}^2\rangle_{\rho}-\langle\hat{O}\rangle^2_{\rho}$ denotes the
variance of the operator $\hat{O}$ with respect to the state $\rho$.
This parameter, $\xi^2_{\textrm{opt}}[\rho_t, \hat{\mathbf{S}}]$, quantifies the  achievable metrological
sensitivity enhancement over the standard quantum limit. %metrology using particle separable states.
Its inverse $\xi^{-2}_{\textrm{opt}}[\rho_t,\hat{\mathbf{S}}]>\kappa$, with $1\leq\kappa\leq (N-1)$,
reveals the multiparticle entanglement of at least $(\kappa+1)$ qubits \cite{Pezze2009}.
	 When the observables are limited to linear collective spin operators $\hat{\mathbf{S}}_{(1)}=\hat{\mathbf{J}}$, the  $\xi^2_{\textrm{opt}}[\rho_t, \hat{\mathbf{J}}]$ reduces to the linear RSP $\xi^2_{\textrm{R}}[\rho_t]$. We can further
	 define the NLSP $\xi_{\textrm{NL}}^2[\rho_t]$ with an  $\hat{\mathbf{S}}$
	 that includes not only linear  but
	  also nonlinear operators. For example,  the second-order NLSP \cite{Gessner2019} corresponds to the $D=9$  linear and quadratic collective spin operators in different directions: %(Fig.~\ref{fig2}A):
$ \hat{\mathbf{S}}_{(2)}=(\hat{J}_x,\hat{J}_y,\hat{J}_z,\hat{J}_x^2,\hat{J}_y^2,\hat{J}_z^2,\hat{J}_{xy}^2,\hat{J}_{yz}^2,\hat{J}_{zx}^2)$,
 where ${\hat{J}_{\beta\gamma}}\equiv{(\hat{J}_\beta+\hat{J}_\gamma)}/{\sqrt{2}}$, with $\beta,\gamma\in\{x,y,z\}$.

\begin{figure}[t]
\centering
\includegraphics[width=0.99\linewidth]{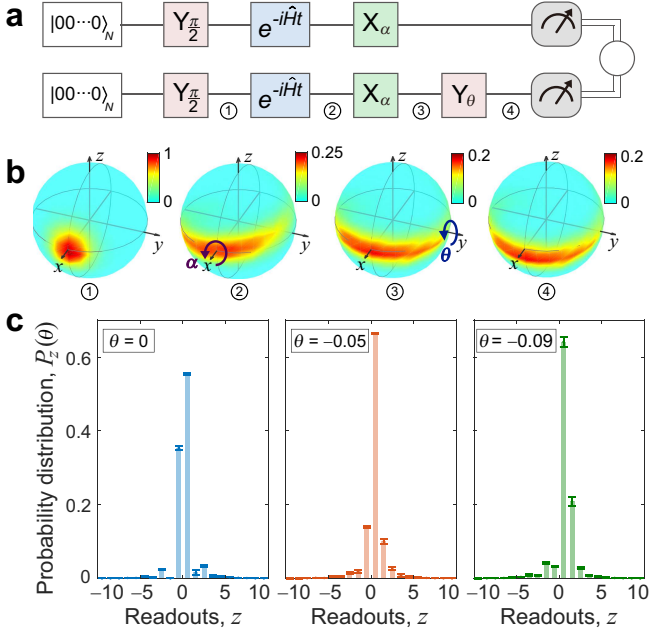}
\caption{\textbf{Experimental procedure for extracting the Fisher information for 19 superconducting qubits.}
\textbf{a}, Schematic of the preparation, nonlinear evolution, optimisation rotation, measurement, and
comparison of the readouts of two states, with and without the collective phase  pulse Y$_\theta$, $\exp(-i\hat{J}_y\theta)$, as in the Ramsey interferometer, for the extraction of the Fisher information.
\textbf{b}, Experimental data of the $Q$ functions, $Q(\theta,\phi)$,  representing the states after \textcircled{\scriptsize 1}  the initial preparation, \textcircled{\scriptsize 2} the nonlinear evolution with  $t=48$~ns, \textcircled{\scriptsize 3} the
collective optimisation rotation  X$_\alpha$, $\exp(-i\hat{J}_x\alpha)$, with $\alpha=-0.288$~rad, and \textcircled{\scriptsize 4} the Ramsey pulse Y$_\theta$, respectively.
\textbf{c}, The probability distributions $P_z(\theta)$ of the measurement observable $\hat{J}_z$ from the single-shot readout of each qubit, for $\theta=0$~rad, $-0.05$~rad, and $-0.09$~rad.
}\label{fig3}
\end{figure}

\begin{figure}[t]
\centering
\includegraphics[width=0.99\linewidth]{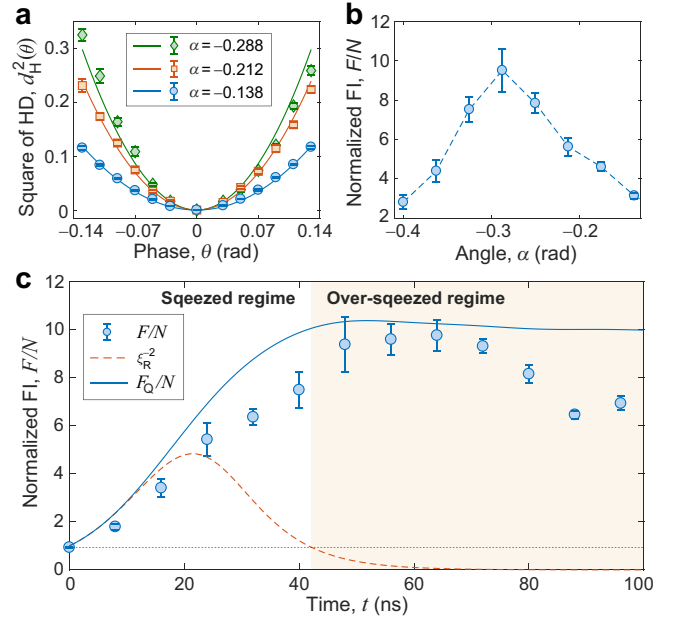}
\caption{\textbf{Quantum enhanced metrology with 19 superconducting qubits.}
\textbf{a}, The squared Hellinger distance (HD) at $t=48$~ns versus the phase $\theta$ in the Ramsey interferometer for different tomography angles $\alpha=-0.288$~rad, $-0.212$~rad, and $-0.138$~rad (along the $x$-axis). The solid lines are for the quadratic curve fitting. \textbf{b},  The normalised Fisher information (FI), $F/N$, extracted
from the squared Hellinger distance versus $\alpha$. The optimal angle at $t=48$~ns is obtained as $\alpha_{\textrm{opt}}=-0.288$~rad.
\textbf{c}, The time evolution of the normalised FI, $F/N$, is compared with the numerical simulations of the inverse RSP (red dashed curve), $\xi^{-2}_{\textrm{R}}$, and the normalised quantum Fisher information (blue solid curve), $F_{\textrm{Q}}/N$, without decoherence. The error bars, indicating the standard deviations of the results, are calculated from about 600,000 repetitive experimental runs in total (see Supplementary Information for details).}\label{fig4}
\end{figure}

With the method  in ref.~\cite{Gessner2019} to optimise measurement observables for quantum metrology,
we first measure the linear RSP  and the NLSP of  $N=10$ qubits during the  nonlinear free evolution.
The optimal metrological squeezing parameter can be obtained via searching the maximum eigenvalue
$\lambda_{\textrm{max}}$ of a $3\times 3$ matrix $\tilde{\mathbb{M}}[\rho,\hat{\mathbf{S}}]$
 as \cite{Gessner2019}
 \begin{align}
\xi_{\textrm{opt}}^2[\rho_t,\hat{\mathbf{S}}]=\frac{N}{\lambda_{\textrm{max}}(\tilde{\mathbb{M}}[\rho_t,\hat{\mathbf{S}}])},
 \end{align}
where $\tilde{\mathbb{M}}$ is the submatrix only containing the first three rows and columns of a $D\times D$ matrix ${\mathbb{M}}$.
The matrix ${\mathbb{M}}$ reads
\begin{align}
\mathbb{M}[\rho_t,\hat{\mathbf{S}}]=\mathbb{C}^T[\rho_t,\hat{\mathbf{S}}]\mathbb{V}^{-1}[\rho_t,\hat{\mathbf{S}}]\mathbb{C}[\rho_t,\hat{\mathbf{S}}],
\end{align}
where $\mathbb{V}[{\rho_t},\hat{\mathbf{S}}]$ is the covariance matrix, with elements
$\mathbb{V}_{ij}[{\rho_t},\hat{\mathbf{S}}]={\langle\{\hat{S}_i,\hat{S}_j\}\rangle_{\rho_t}}/{2}-\langle\hat{S}_i\rangle_{\rho_t}\langle\hat{S}_j\rangle_{\rho_t}$,
and $\mathbb{C}[\rho_t,\hat{\mathbf{S}}]$ is the real-valued skew-symmetric commutator matrix, with elements
$\mathbb{C}_{ij}[\rho_t,\hat{\mathbf{S}}]=-i\langle[\hat{S}_i,\hat{S}_j]\rangle_{\rho_t}$.
For  simplicity, we merely select seven collective spin operators (see Methods),
$\hat{\mathbf{S}}_{\textrm{exp}}=(\hat{J}_x,\hat{J}_y,\hat{J}_z,\hat{J}_x^2,\hat{J}_y^2,\hat{J}_{xy}^2,\hat{J}_{zx}^2)$,
and obtain the time evolution of the NLSP via measuring each element of
$\mathbb{V}[\rho_t,\hat{\mathbf{S}}_{\textrm{exp}}]$ and $\mathbb{C}[\rho_t,\hat{\mathbf{S}}_{\textrm{exp}}]$ with
simultaneous single-shot readouts of  10 qubits in different directions (see Supplementary Information for more details).
The RSP can be simply given by considering the submatrices
$\tilde{\mathbb{V}}$ and $\tilde{\mathbb{C}}$, only containing the first three rows and columns of $\mathbb{V}$ and
$\mathbb{C}$, respectively. At  $t=34$~ns and 50~ns, the experimental
data of matrices $\mathbb{M}[\rho_t,\hat{\mathbf{J}}]$ and $\mathbb{M}[\rho_t,\hat{\mathbf{S}}_{\textrm{exp}}]$
for the RSP and the NLSP, respectively, are compared with the numerical predictions in Fig.~\ref{fig2}a,b.
The time evolutions of the inverse RSP, $\xi_{\textrm{R}}^{-2}$, and the inverse NLSP, $\xi_{\textrm{NL}}^{-2}$,
are shown in Fig.~\ref{fig2}d, which are compared with the normalised Fisher information, $F/N$.
Our results, verifying the
hierarchical relationship, $\xi_{\textrm{R}}^{-2}\leq \xi_{\textrm{NL}}^{-2}$,
demonstrate that the NLSP, generalising and  improving the RSP with  additional quadratic operators,
helps to capture a larger set of metrologically useful entangled states.
Especially, in the over-squeezed regime (e.g., $t=82$~ns), the NLSP and the Fisher information identify the
multiparticle entangled state with an obvious non-Gaussian distribution in phase space (Fig.~\ref{fig2}c),
which \emph{cannot} be characterised by the linear RSP.
Therefore,  the NLSP, measured with single-shot readouts of collective spin operators, is efficient to capture
the entanglement of the non-Gaussian state without the need of quantum state tomography.

Furthermore, the maximal Fisher information, $F_{\textrm{opt}}$, which quantifies the achievable metrological
sensitivity with the optimal linear observable and  linear generator,
 gives an upper bound to the inverse of the optimal metrological squeezing parameter
$F_{\textrm{opt}}/N\geq\xi_{\textrm{opt}}^{-2}[\rho_t,\hat{\mathbf{S}}]$ \cite{Gessner2019}. To demonstrate the
metrological performance of our superconducting qubits, we experimentally detect the Fisher information
by comparing the measurement statistics of the time evolved states $\tilde{\rho}_t(0)$ and $\tilde{\rho}_t(\theta)$
with and without a small rotation with the generator $\hat{J}_{y}$ after the optimisation rotation along the $x$-axis (Fig.~\ref{fig3}a).
%, $\hat{J}_{\bm{n}_{\textrm{opt}}}=\exp(i\hat{J}_{x}\alpha_{\textrm{opt}})\hat{J}_{y}\exp(-i\hat{J}_{x}\alpha_{\textrm{opt}})$.
For a small
$\theta$ and sufficiently large number of experimental realisations, the Fisher information can be extracted as the coefficient
of the quadratic term from a polynomial fit to  the square of the Hellinger distance \cite{Pezze2018}: (Fig.~\ref{fig4}a)
\begin{align}
	d^2_{\textrm{H}}(\theta)=\frac{F}{8}\theta^2+\mathcal{O}(\theta^3),
\end{align}
where $d^2_{\textrm{H}}(\theta)\equiv 1-\sum_{z}\sqrt{P_z(0)P_z(\theta)}$, and the sum
is  the Bhattacharyya coefficient with probability distributions $P_z(0)$ and $P_z(\theta)$
of the observable $\hat{J}_z$  for states  $\tilde{\rho}_t(0)$ and $\tilde{\rho}_t(\theta)$,
respectively (Fig.~\ref{fig3}c). In Fig.~\ref{fig2}d, we  show with $N=10$ qubits that
the Fisher information reveals larger multiparticle entanglement, though
the RSP and the NLSP increase at long evolution times (e.g., $F/N\geq\xi_{\textrm{NL}}^{-2}\geq\xi_{\textrm{R}}^{-2}$,
with $t\geq66$~ns). The maximum normalised Fisher information, $F/N=5.13\pm0.32$
($7.10^{+0.26}_{-0.28}$~dB),  is detected at $t=66$~ns for $N=10$ qubits. For $N=19$ qubits, we measure the
Fisher information during the nonlinear time evolution as shown in Fig.~\ref{fig4}c. At $t=64$~ns, we observe
the maximum metrological gain $F/N=9.75\pm0.64$ ($9.89^{+0.28}_{-0.29}$~dB), benefiting from the
multiparticle entanglement of non-Gaussian states in the over-squeezed regime.

We have demonstrated the characterisation of multiparticle entangled states of superconducting qubits utilising different
concepts of entanglement witnesses, including the RSP, the NLSP, and the Fisher information.
With 19 qubits, we have obtained a larger quantum metrological gain over the classical metrology than those
obtained in many other platforms with much larger number of particles, representing the potential
capability of showing quantum advantages in practical quantum metrology with interconnected superconducting
qubits. Owing to the high-fidelity controls and  individually addressable single-shot readouts of qubits
 with long decoherence times, our system is also promising for realising different quantum algorithms including
 two-axis-twisting spin squeezing \cite{KITAGAWA1993} and variational quantum simulations \cite{Kokail2019}.\\

\noindent\textbf{Acknowledgements}
This work was supported by
the National Key R\&D Program of China (Grant Nos.~2016YFA0302104, 2016YFA0300600),
the National Natural Science Foundation of China (Grant Nos.~11774406, 11934018, 11904393, 92065114),
the Strategic Priority Research Program of Chinese Academy of Sciences (Grant No.~XDB28000000),
%Japan Society for the Promotion of Science (JSPS) Postdoctoral Fellowship (Grant No.~P19326),
%JSPS KAKENHI (Grant No.~JP19F19326),
the NTT Research,
the Army Research Office (ARO) (Grant No.~W911NF-18-1-0358),
the Japan Science and Technology Agency (JST)
(via the Q-LEAP program and the CREST Grant No.~JPMJCR1676),
the Japan Society for the Promotion of Science (JSPS)
(via the Postdoctoral Fellowship Grant No.~P19326, the KAKENHI Grant Nos.~JP19F19326, JP20H00134
and the JSPS-RFBR Grant No.~JPJSBP120194828),
the Asian Office of Aerospace Research and Development (AOARD) (Grant No.~FA2386-20-1-4069),
and the Foundational Questions Institute Fund (FQXi) (Grant No.~FQXi-IAF19-06).
\\

\noindent\textbf{Author contributions}
H.F., Y.-R.Z., and H.W. conceived the experiment. K.X. carried out the measurements.
K.X. and Z.-H.S. analysed the experimental data. H.L. and D.Z. fabricated the device.
Y.-R.Z., Z.-H.S. and K.X. designed the experiment and performed theoretical analysis.
Y.-R.Z., K.X., Z.-H.S., F.N., H.W., and H.F. cowrote the manuscript.
All authors contributed to the experimental setup, discussions of the results, and development of the manuscript.\\

\noindent\textbf{Competing interests}
The authors declare no competing interests.\\

\noindent\textbf{Data availability}
Data used in this work is available on reasonable request.\\

\noindent\textbf{Code availability}
Code used in this work is available on reasonable request.

\clearpage

\renewcommand{\figurename}{Extended Data Fig.}
\renewcommand{\tablename}{Extended Data Table}

\section*{Methods}
\subsection*{Experimental device}\label{sec:2}

The device contains 20 superconducting qubits (q$_j$ with $j$ varied from 1 to 20), which are fully connected through a common
resonator bus R.
In our experiments, we use 19 of them, as one qubit (q$_7$) suffers from a strong interaction with a
two-level system near its working point. The qubit characteristics can be found in
ref.~\cite{Song2019}, which is the same device used in this experiment.
Extended Data Table~\ref{table1} lists the latest device information for the participating qubits, obtained
during this experiment. The full Hamiltonian of our quantum processor can be written as
\begin{align}
	\hat{H}_0= &\;\omega_r \hat{a}^{\dag}\hat{a} + \sum_{j=1}^{19}\omega_{j}\hat{\sigma}_{j}^+\hat{\sigma}_{j}^-
	+\sum_{j=1}^{19}g_{j}(\hat{\sigma}_{j}^{+}\hat{a}+\hat{\sigma}_{j}^{-}\hat{a}^{\dag})\nonumber\\
	&+\sum_{i<j}^{}\chi^{c}_{ij}(\hat{\sigma}^{+}_{i}\hat{\sigma}^{-}_{j}+\hat{\sigma}^{+}_{j}\hat{\sigma}^{-}_{i}),
	\label{Seq:no1}
\end{align}
where $\omega_j/2\pi$ denotes the resonant frequency of q$_j$ (individually tuneable from
3~GHz to 5.5~GHz). The frequency of the common resonator bus R, represented by
$\omega_r/2\pi$, is fixed at about 5.51~GHz.
Each qubit q$_j$ is capacitively coupled to R, with magnitude, $g_j/2\pi$, listed in
Extended Data Table~\ref{table1}. Note that except for the dominant qubit-resonator interaction, there exist
small direct couplings, $\chi^{c}_{ij}/2\pi$, between qubits in the system. In our experiments, by
equally detuning the frequencies of all qubits far away from that of the resonator, we can
realise the resonator-induced super-exchange interaction with a magnitude of $g_i g_j/(2\pi\Delta)$
(with $\Delta=\omega_i-\omega_r=\omega_j-\omega_r$, and $\vert\Delta\vert\gg g_i,g_j$)
between \emph{any} two qubits. The Hamiltonian can be further written as
\begin{equation}
	\hat{H}=\sum_{i<j}^{}(g_i g_j/\Delta+\chi^{c}_{ij})(\hat{\sigma}^{+}_{i}\hat{\sigma}^{-}_{j}+\hat{\sigma}^{+}_{j}\hat{\sigma}^{-}_{i}).
	\label{Seq:no2}
\end{equation}
The qubit-qubit coupling strengths
\begin{align}
	\chi_{ij}\equiv g_i g_j/\Delta+\chi^{c}_{ij},
\end{align}
 which can be experimentally estimated by the energy swapping process between q$_i$ and q$_j$
 (see Supplementary materials of ref.~\cite{Song2017}), are shown in
Extended Data Fig.~\ref{coupling matrix}, with $\Delta/2\pi\simeq-580$~MHz in this experiment.

For the 10-qubit experiment, we choose q$_6$, q$_{9}$, q$_{10}$, q$_{11}$, q$_{12}$, q$_{13}$,
q$_{14}$, q$_{17}$, q$_{18}$, and q$_{20}$. For the 19-qubit experiment, we choose 19 qubits
except for q$_7$. In the main text, for convenience, the 19 qubits in order \{q$_6$, q$_{9}$,
q$_{10}$, q$_{11}$, q$_{12}$, q$_{13}$, q$_{14}$, q$_{17}$, q$_{18}$, q$_{20}$, q$_1$, q$_{2}$,
q$_{3}$, q$_{4}$, q$_{5}$, q$_{6}$, q$_{15}$, q$_{16}$, q$_{19}$\} are relabelled as \{Q$_j$\} with $j=1,2,\cdots,19$.

\subsection*{Phase calibration}
In our experiments, the nonlinear evolution $\exp({-i\hat{H}t})$ is realised by equally detuning
all the qubits from their idle points, $\omega_j/2\pi$, to the interacting point, $\omega_I/2\pi$, by
applying a rectangular pulse to each qubit. This operation will accumulate some dynamical
phase, which needs to be cancelled via applying rotation pulses after the rectangular pulses.
In theory, the dynamical phase can be estimated as $2\pi \delta\! f\times t$, where
$\delta \!f=(\omega_j-\omega_I)/2\pi$. However, the imperfections of rectangular pulses,
such as the imperfect rising and falling edges, will cause an additional phase shift from the
theoretical calculations, which also needs to be experimentally calibrated. Extended Data Figure~\ref{phaseCali}
shows the pulse sequence and the results of our phase calibration method, taking q$_1$ as
an example. q$_1$ is tuned to the interacting point, while other qubits are arranged at their
frequency points $\omega_j^o/2\pi$ far away from $\omega_I/2\pi$. To minimise the Z-crosstalk
effects of other qubits to q$_j$ when being tuned away, $\omega_j^o/2\pi$ are selected to have
an equal Z-crosstalk effect on q$_1$, compared to the case when all qubits are tuned to
$\omega_I/2\pi$, as can be estimated by the measured Z-crosstalk matrix $M_Z$. We monitor
the $\vert 1\rangle$ state probability $P(\phi,t)$ as a function of both the time $t$ and the phase difference $(\phi-2\pi\delta\! f\times t)$. For each time $t$, we perform a
cosine fit to $P(\phi,t)$ as a function of $\phi$ to extract the phase shift $\phi_+^c$, caused
by the imperfect rectangular pulses. To further reduce the Z-crosstalk effects and the ac-stark
shift effects due to imperfect decoupling of other qubits to q$_1$ when being tuned away, we perform this calibration process again with a little difference. The frequencies of other qubits are arranged at $(2\omega_I-\omega_j^o)/2\pi$, a symmetric position relative to $\omega_I/2\pi$. Again, we obtain the phase shift $\phi_-^c$. The final phase shift, used to cancel the dynamical phase, is $(\phi_+^c+\phi_-^c)/2$, as shown by the red curves in Extended Data Fig.~\ref{phaseCali}c.

\subsection*{Efficient detection of second-order nonlinear squeezing parameter with seven operators}
In our experiments, to obtain the nonlinear squeezing parameter, we select seven collective spin operators,
\begin{align}
\hat{\mathbf{S}}_{\textrm{exp}}=(\hat{J}_x,\hat{J}_y,\hat{J}_z,\hat{J}_x^2,\hat{J}_y^2,\hat{J}_{xy}^2,\hat{J}_{yz}^2),\label{seven}
\end{align}
instead of the full family of the collective spin operators
\begin{align}
\hat{\mathbf{S}}_{(2)}=(\hat{J}_x,\hat{J}_y,\hat{J}_z,\hat{J}_x^2,\hat{J}_y^2,\hat{J}_z^2,\hat{J}_{xy}^2,\hat{J}_{yz}^2,\hat{J}_{zx}^2),\label{nine}
\end{align}
for the second-order squeezing parameter \cite{Gessner2019}. As shown in Fig.~{\color{magenta}2}d in the main text, we
monitor the evolution of the nonlinear squeezing parameter via
measuring each element of $\mathbb{V}[\rho_t,\hat{\mathbf{S}}_{\textrm{exp}}]$ and
$\mathbb{C}[\rho_t,\hat{\mathbf{S}}_{\textrm{exp}}]$ (submatrices of
$\mathbb{V}[\rho_t,\hat{\mathbf{S}}_{(2)}]$ and $\mathbb{C}[\rho_t,\hat{\mathbf{S}}_{(2)}])$
with simultaneous single-shot readouts of 10 qubits in different directions (see Supplementary Information).
The numerical simulations of the inverse nonlinear squeezing parameters,
$\xi^{-2}_{\textrm{NL}}[\rho_t,\hat{\mathbf{S}}_{(2)}]$ and
$\xi^{-2}_{\textrm{NL}}[\rho_t,\hat{\mathbf{S}}_{\textrm{exp}}]$ with respect to
$\hat{\mathbf{S}}_{(2)}$ and $\hat{\mathbf{S}}_{\textrm{exp}}$, respectively, are compared
in Extended Data Fig.~\ref{fs6}. It is shown that the nonlinear squeezing parameter
$\xi^{-2}_{\textrm{NL}}[\rho_t,\hat{\mathbf{S}}_{\textrm{exp}}]$ with 7 selected
collective spin operators is
very close to the second-order nonlinear squeezing parameter $\xi^{2}_{\textrm{NL}}[\rho_t,\hat{\mathbf{S}}_{\textrm{exp}}]$ for
$t\lesssim76$~ns, and moreover,  its minimum value is close to that of the
$\xi^{2}_{\textrm{NL}}[\rho_t,\hat{\mathbf{S}}_{(2)}]$.
Therefore, when choosing these 7 collective spin operators,
we can efficiently detect the second-order nonlinear squeezing parameter with fewer observables to
be measured. At $t=2$~ns, $34$~ns, and 50~ns, the experimental results of matrices $\mathbb{C}[\rho_t,\hat{\mathbf{S}}_{\textrm{exp}}]$, $\mathbb{V}[\rho_t,\hat{\mathbf{S}}_{\textrm{exp}}]$, and $\mathbb{M}[\rho_t,\hat{\mathbf{S}}_{\textrm{exp}}]$
are compared with numerical simulations in Extended Data Figs.~\ref{fs3}, \ref{fs4}, and \ref{fs5}, respectively.

In addition, the method to optimise the squeezing parameter, based on searching for the largest
eigenvalue of the matrix $\mathbb{M}$ \cite{Gessner2019}, requires a large number of trials of single-shot
readouts for 19 observables (see Supplementary Information) to obtain a reliable value of the second-order squeezing parameter.
Thus, choosing 7 collective spin operators instead of 9 operators in our experiments can significantly reduce
the number of readouts, and almost detect the large sensitivity enhancement for quantum metrology, which
is characterised by the second-order nonlinear squeezing parameter with
9  operators.

\subsection*{Numerical details}
Numerical computations are performed using the \textsc{QuTiP} \cite{Johansson:2012aa,Johansson:2013aa}
(the quantum toolbox in \textsc{Python}) and \textsc{NumPy}.
The time evolutions of the system with a Hamiltonian [Eq.~({\color{magenta}1}) in the main text]  are numerically
simulated using \textsc{QuTiP}'s
master equation solver \textsf{mesolve}, where the parameters in Extended Data Fig.~\ref{coupling matrix} are used.
Because the evolution time is much shorter than the qubits' energy relaxation time and dephasing time
$t\ll T_1, T_2$, we neglect the effect of decoherence in simulations.

\setcounter{figure}{0}
\begin{figure*}[b]
	\centering
	\includegraphics[width=0.98\linewidth,clip=True]{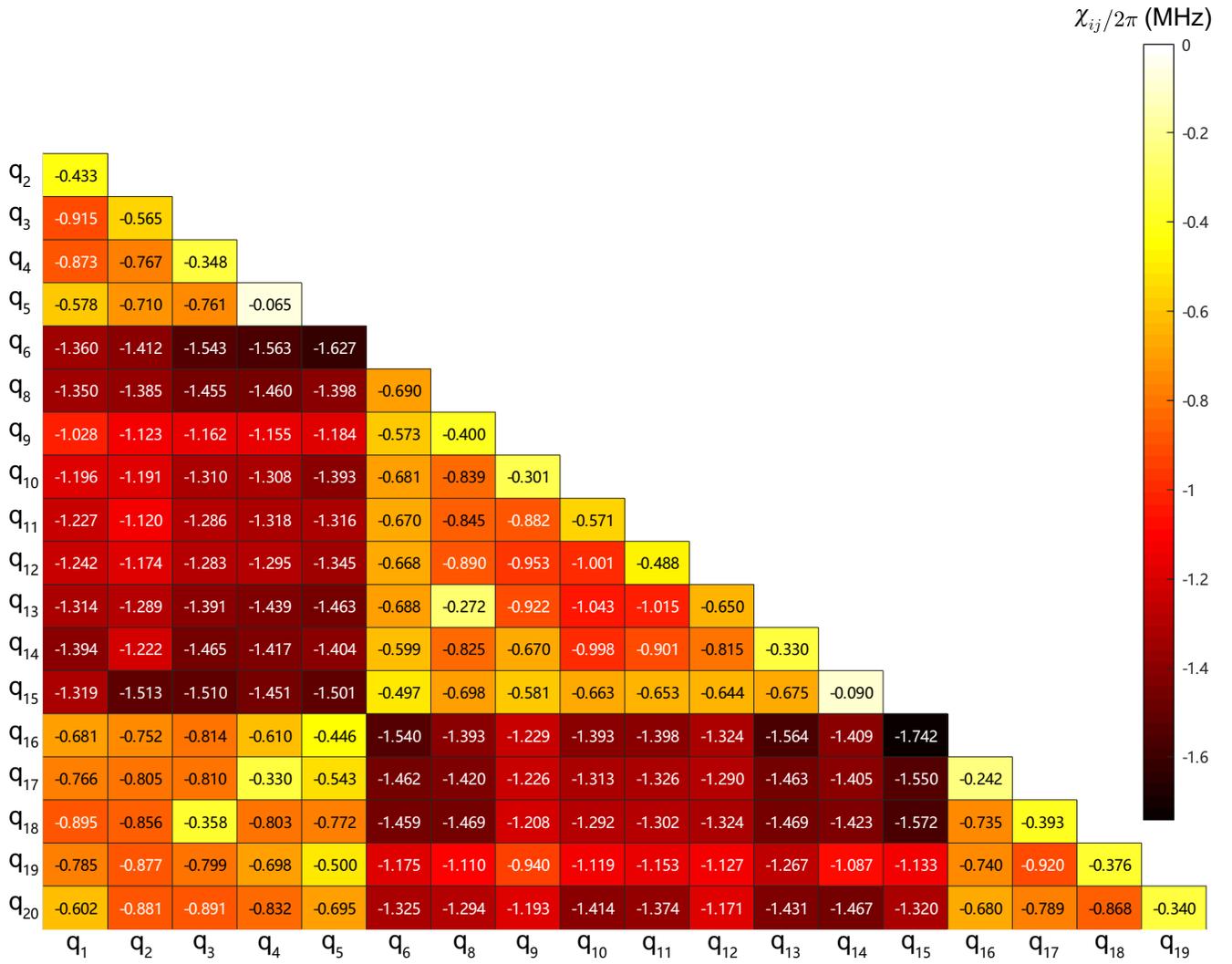}
	\caption{\textbf{Coupling matrix.} Plotted is the coupling strength $\chi_{ij}/2\pi$ between q$_i$ and q$_j$ in the quantum processor, which is measured by the energy swapping process, where q$_i$ and q$_j$ are equally tuned at the working point $\omega_I$ to interact for a specific time \cite{Song2017}.
		\label{coupling matrix}}
\end{figure*}

\begin{table*}[b]
	\centering
	\begin{tabular}{c|cccccccccccc}
		%\centering
		\hline
		\hline
		~~~~~~~~~~~~~~~&~$\omega_{j}/2\pi$ (GHz) ~&~~~$T_{1,j}$ ($\mu$s) ~~~&~$g_j/2\pi$ (MHz)~&~$\omega_j^r/2\pi$ (GHz) ~&~$\omega_{j}^{m}/2\pi$ (GHz) &~~~~~~~$F_{0,j}$~~~~~~~&~~~~~~~$F_{1,j}$~~~~~~~\\
%		&&&&&(GHz)&(GHz)&&\\
		\hline
		q$_1$      & 4.350 & $\approx20$  & 27.6 & 6.768 & 4.460 & 0.977 & 0.921\\
		q$_2$      & 4.390 & $\approx26$  & 27.4 & 6.741 & 4.310 & 0.986 & 0.879\\
		q$_3$      & 4.275 & $\approx27$  & 29.1 & 6.707 & 4.355 & 0.975 & 0.912\\
		q$_4$      & 4.300 & $\approx26$  & 27.6 & 6.676 & 4.440 & 0.989 & 0.918\\
		q$_5$      & 4.245 & $\approx26$  & 26.5 & 6.649 & 4.260 & 0.975 & 0.909\\
		q$_6$      & 5.081 & $\approx27$  & 29.2 & 6.612 & 4.805 & 0.975 & 0.925\\
		q$_8$      & 4.215 & $\approx26$  & 30.1 & 6.558 & 4.285 & 0.987 & 0.906\\
		q$_9$      & 5.120 & $\approx23$  & 24.1 & 6.552 & 5.070 & 0.989 & 0.926\\
		q$_{10}$  & 5.160 & $\approx30$ & 27.7 & 6.514 & 5.290 & 0.995 & 0.903\\
		q$_{11}$  & 5.290 & $\approx24$  & 27.3 & 6.525 & 5.170 & 0.994 & 0.897\\
		q$_{12}$  & 5.215 & $\approx35$  & 26.9 & 6.550 & 5.210 & 0.981 & 0.920\\
		q$_{13}$  & 4.945 & $\approx26$  & 29.1 & 6.568 & 4.895 & 0.980 & 0.916\\
		q$_{14}$     & 5.250 & $\approx41$  & 27.4 & 6.598 & 5.250 & 0.983 & 0.896\\
		q$_{15}$  & 4.895 & $\approx31$  & 26.3 & 6.641 & 4.235 & 0.978 & 0.913\\
		q$_{16}$  & 4.325 & $\approx25$  & 26.5 & 6.660 & 4.850 & 0.987 & 0.934\\
		q$_{17}$     & 4.735 & $\approx36$  & 27.3 & 6.686 & 4.578 & 0.984 & 0.942\\
		q$_{18}$  & 4.815 & $\approx38$  & 29.0 & 6.713 & 4.770 & 0.982 & 0.912\\
		q$_{19}$  & 4.425 & $\approx35$  & 24.6 & 6.788 & 4.385 & 0.98 & 0.900\\
		q$_{20}$  & 4.855 & $\approx30$  & 27.5 & 6.759 & 5.115 & 0.985 & 0.918\\
		\hline
		\hline
	\end{tabular}
	\caption{\label{table1} \textbf{Qubit characteristics.} $\omega_{j}/2\pi$ is the idle frequency of q$_j$, where single-qubit rotation pulses are applied. $T_{1,j}$ is the energy relaxation time of q$_j$, which is the typical value across a wide frequency range. $g_j/2\pi$ denotes the coupling strength between q$_j$ and the resonator bus R. $\omega_j^r/2\pi$ is the resonant frequency of q$_j$'s readout resonator. $\omega_{j}^{m}/2\pi$ is the resonant frequency of q$_j$ at the beginning of the measurement process, when its readout resonator is pumped with microwave pulse. $F_{0,j}$ ($F_{1,j}$) is the probability of detecting q$_j$ in $\vert 0\rangle$ ($\vert 1\rangle$) state, when it is prepared in the $\vert 0\rangle$ ($\vert 1\rangle$) state.}
\end{table*}

\begin{figure*}[t]
	\centering
	\includegraphics[width=0.55\linewidth,clip=True]{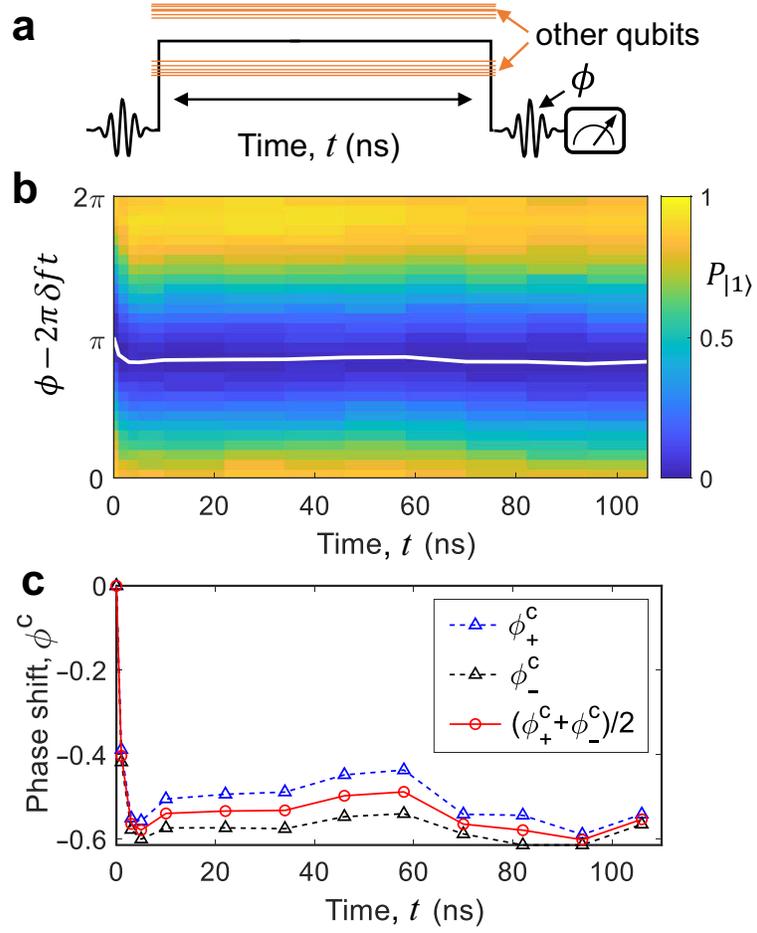}
	\caption{\textbf{Dynamical phase calibration.} \textbf{a}, Experimental results and sequence of phase calibration for q$_1$. \textbf{b}, The phase shift $\phi^c$ obtained by fitting the results in (\textbf{a}) as a function of the interaction time $t$.
  \label{phaseCali}}
\end{figure*}

\begin{figure*}[t]
\centering
\includegraphics[width=0.6\linewidth,clip=True]{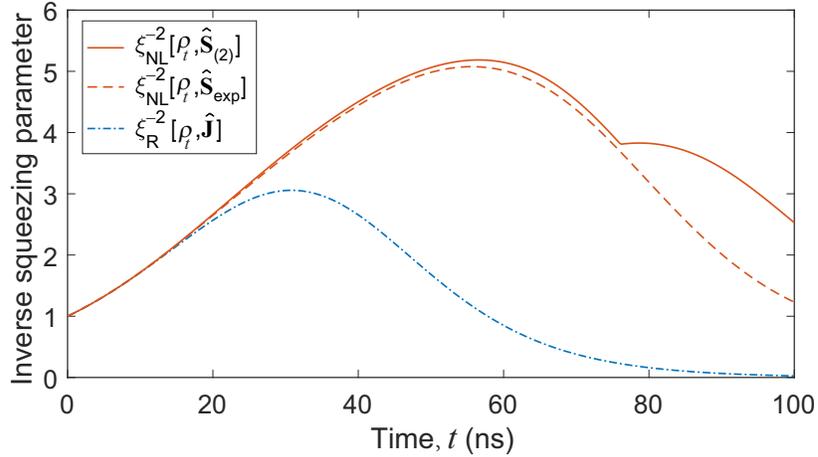}
\caption{\textbf{Efficient detection of second-order nonlinear squeezing parameter with
seven operators.} Numerical simulations of the evolutions of the inverse Ramsey squeezing parameter,
$\xi^{-2}_{\textrm{R}}[\rho_t,\hat{\mathbf{J}}]$, with a family of 3 collective spin operators $\hat{\mathbf{J}}=(\hat{J}_x,\hat{J}_y,\hat{J}_z)$, the
inverse second-order nonlinear squeezing parameter, $\xi^{-2}_{\textrm{NL}}[\rho_t,\hat{\mathbf{S}}_{(2)}]$,
with 9 operators in Eq.~(\ref{nine}),
and the inverse nonlinear squeezing parameter, $\xi^{-2}_{\textrm{NL}}[\rho_t,\hat{\mathbf{S}}_{\textrm{exp}}]$
with 7 operators in Eq.~(\ref{seven}).
}\label{fs6}
\end{figure*}

\begin{figure*}[t]
\centering
\includegraphics[width=0.7\linewidth]{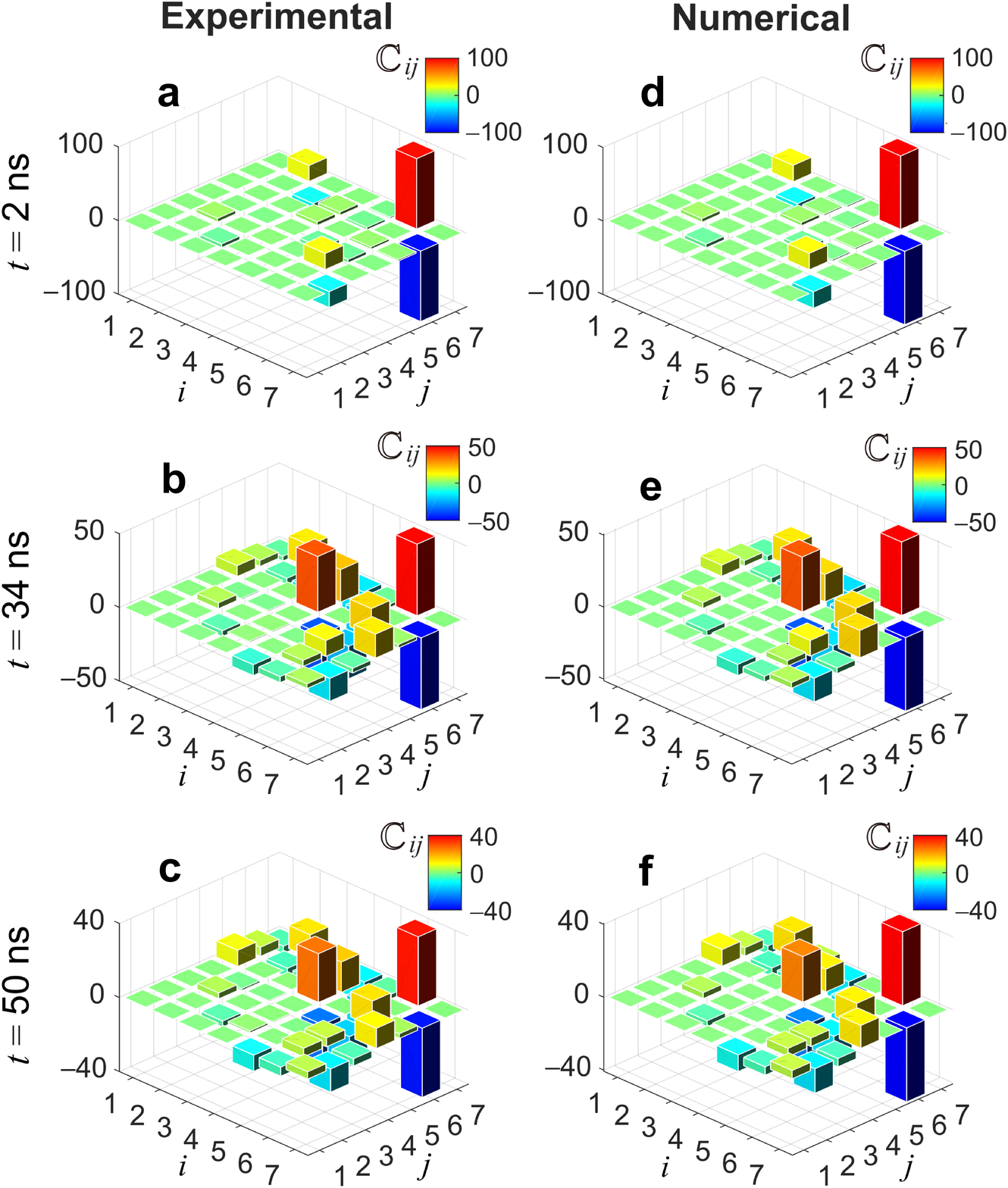}
\caption{\textbf{Data of the $\mathbb{C}$ matrix.} \textbf{a--f},
Matrix $\mathbb{C}[\rho_t,\hat{\mathbf{S}}_{\textrm{exp}}]$ experimentally measured at (\textbf{a}) $t=2$~ns, (\textbf{b}) $t=34$~ns, and (\textbf{c}) $t=50$~ns,
compared with the numerical simulations (\textbf{d}), (\textbf{e}), and (\textbf{f}).
}\label{fs3}
\end{figure*}

\begin{figure*}[t]
\centering
\includegraphics[width=0.7\linewidth]{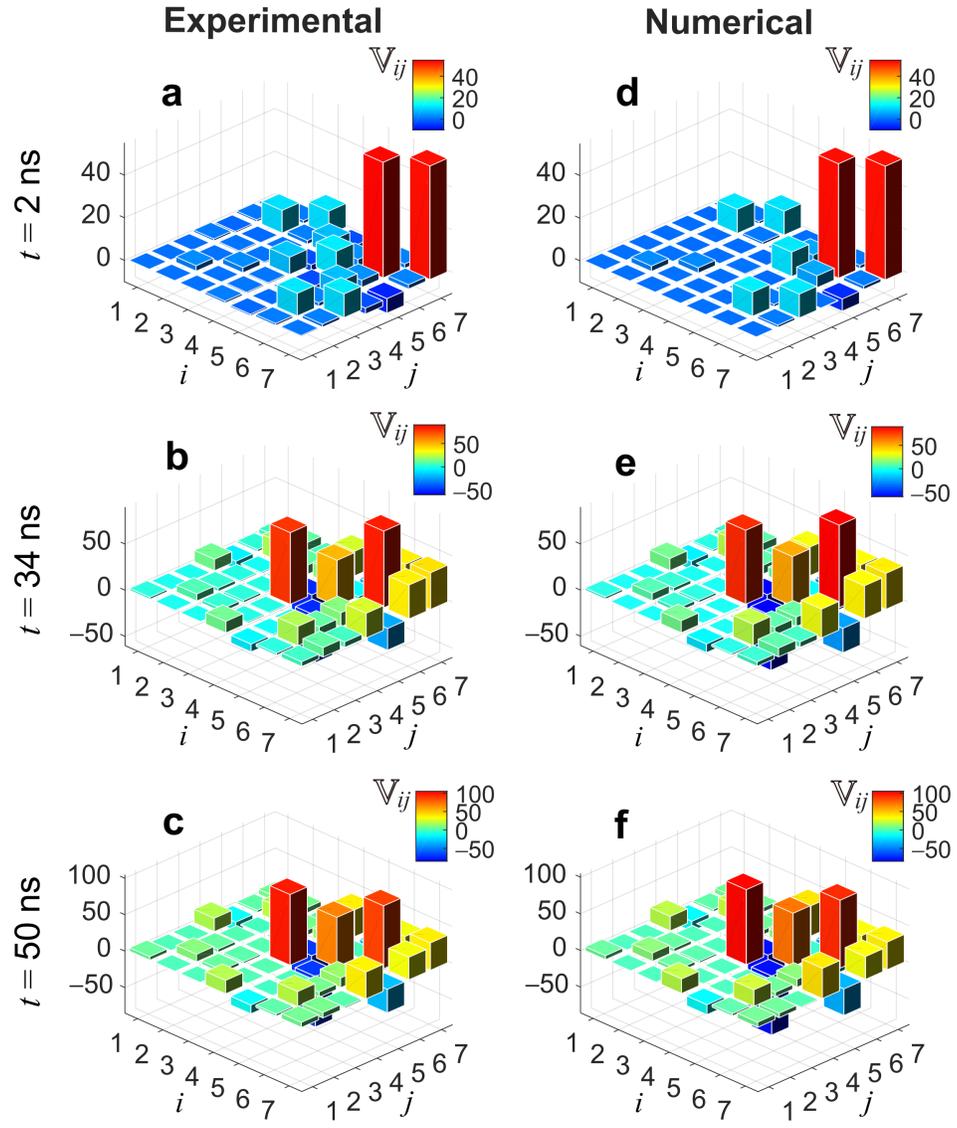}
\caption{\textbf{Data of the $\mathbb{V}$ matrix.} \textbf{a--f},
Matrix $\mathbb{V}[\rho_t,\hat{\mathbf{S}}_{\textrm{exp}}]$ experimentally measured at (\textbf{a}) $t=2$~ns, (\textbf{b}) $t=34$~ns, and (\textbf{c}) $t=50$~ns,
compared with the numerical simulations (\textbf{d}), (\textbf{e}), and (\textbf{f}).
}\label{fs4}
\end{figure*}

\begin{figure*}[t]
\centering
\includegraphics[width=0.7\linewidth]{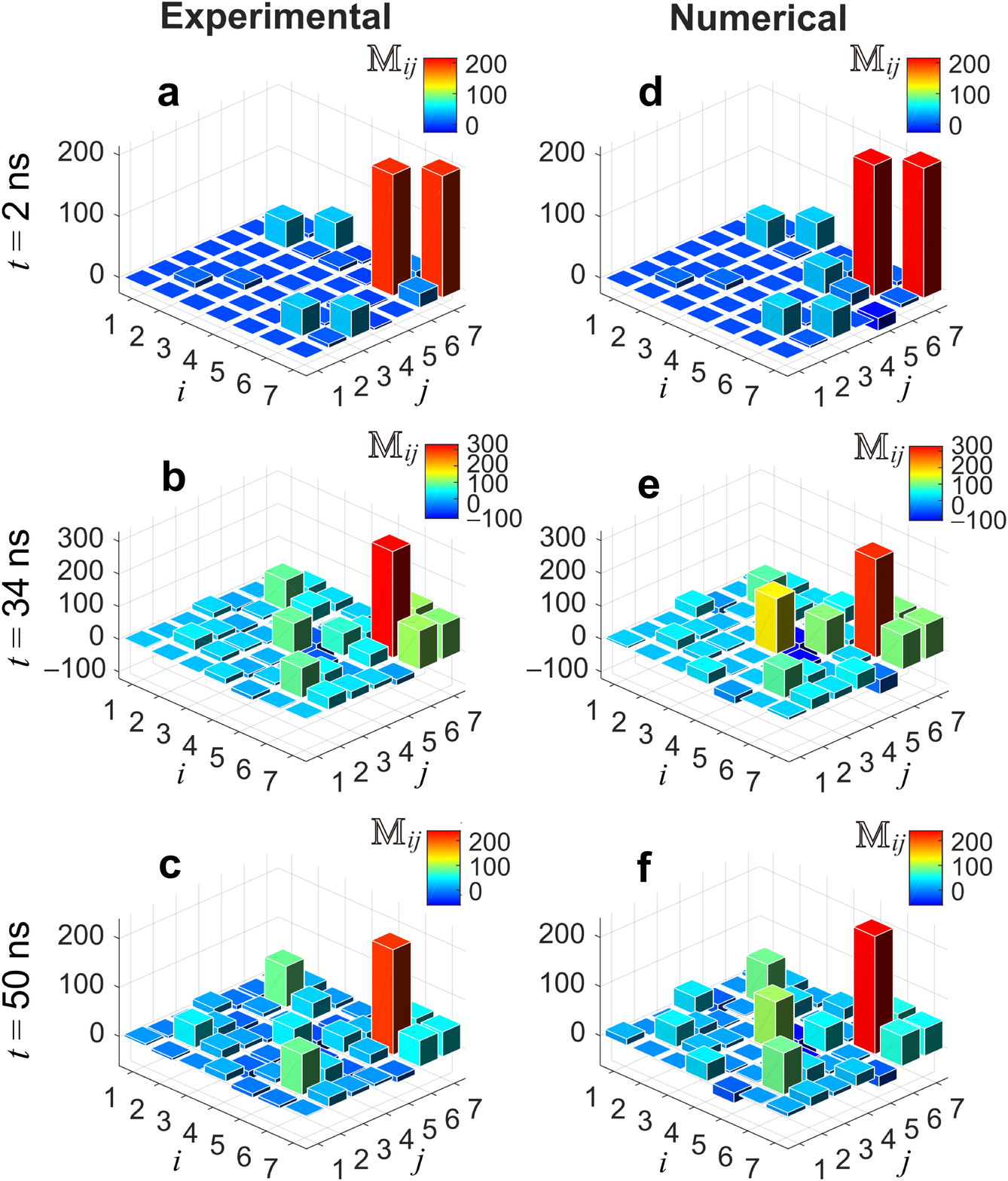}
\caption{\textbf{Data of the $\mathbb{M}$ matrix.} \textbf{a--f},
Matrix $\mathbb{M}[\rho_t,\hat{\mathbf{S}}_{\textrm{exp}}]$ experimentally measured at (\textbf{a}) $t=2$~ns, (\textbf{b}) $t=34$~ns, and (\textbf{c}) $t=50$~ns,
compared with the numerical simulations (\textbf{d}), (\textbf{e}), and (\textbf{f}).
}\label{fs5}
\end{figure*}

\end{document}

\title{Supplementary Information for:  \\%Measuring nonlinear squeezing parameter and approaching effective two-axis twisting spin squeezing with interconnected superconducting qubits\\
%Observing nonlinear squeezing and entanglement of non-Gaussian states with interconnected superconducting qubits\\
Metrological characterisation of non-Gaussian  entangled states of superconducting qubits\\
%\emph{Characterizing multiparticle non-Gaussian entangled states with  superconducting qubits}\\
%\emph{Observing nonlinear squeezing and entanglement of non-Gaussian states with interconnected superconducting qubits}
%Observing spin squeezing with interconnected superconducting qubits
}
\author{Kai Xu}
\thanks{These authors contributed equally to this work.}
\affiliation{Institute of Physics, Chinese Academy of Sciences, Beijing 100190, China}
\author{Yu-Ran Zhang}
\thanks{These authors contributed equally to this work.}
    %\email{yuran.zhang@riken.jp}
    %\affiliation{Beijing Computational Science Research Center, Beijing 100193, China}
	\affiliation{Theoretical Quantum Physics Laboratory, RIKEN Cluster for Pioneering Research, Wako-shi, Saitama 351-0198, Japan}
  \author{Zheng-Hang Sun}
  \thanks{These authors contributed equally to this work.}
  \affiliation{Institute of Physics, Chinese Academy of Sciences, Beijing 100190, China}

  \author{Hekang Li}
  \affiliation{Institute of Physics, Chinese Academy of Sciences, Beijing 100190, China}

  \author{Pengtao Song}
  \affiliation{Institute of Physics, Chinese Academy of Sciences, Beijing 100190, China}

  \author{Zhongcheng Xiang}
\affiliation{Institute of Physics, Chinese Academy of Sciences, Beijing 100190, China}

  \author{Kaixuan Huang}
  \affiliation{Institute of Physics, Chinese Academy of Sciences, Beijing 100190, China}

  \author{Hao Li}
  \affiliation{Institute of Physics, Chinese Academy of Sciences, Beijing 100190, China}

  \author{Yun-Hao Shi}
  \affiliation{Institute of Physics, Chinese Academy of Sciences, Beijing 100190, China}

  \author{Chi-Tong Chen}
  \affiliation{Institute of Physics, Chinese Academy of Sciences, Beijing 100190, China}
  \author{Xiaohui Song}
    %\email{dzheng@iphy.ac.cn}
  \affiliation{Institute of Physics, Chinese Academy of Sciences, Beijing 100190, China}
  \author{Dongning Zheng}
    %\email{dzheng@iphy.ac.cn}
  %\email{hfan@iphy.ac.cn}
  \affiliation{Institute of Physics, Chinese Academy of Sciences, Beijing 100190, China}
  %\affiliation{CAS Centre for Excellence in Topological Quantum Computation, University of Chinese Academy of Sciences, Beijing 100190, China}
	\author{Franco Nori}
	\email{fnori@riken.jp}
		\affiliation{Theoretical Quantum Physics Laboratory, RIKEN Cluster for Pioneering Research, Wako-shi, Saitama 351-0198, Japan}
	  	\affiliation{Physics Department, University of Michigan, Ann Arbor, Michigan 48109-1040, USA}
  \author{H. Wang}
  \email{hhwang@zju.edu.cn}
  \affiliation{Interdisciplinary Centre for Quantum Information, State
Key Laboratory of Modern Optical Instrumentation, and
Zhejiang Province Key Laboratory of Quantum Technology
and Device, Department of Physics, Zhejiang University,
Hangzhou 310027, China}
\author{Heng Fan}
\email{hfan@iphy.ac.cn}
\affiliation{Institute of Physics, Chinese Academy of Sciences, Beijing 100190, China}
\affiliation{CAS Centre for Excellence in Topological Quantum Computation, UCAS, Beijing 100190, China}


\maketitle
\tableofcontents

\renewcommand{\theequation}{S\arabic{equation}}
 %redefine the command that creates the equation no.
\setcounter{equation}{0}  % reset counter

\renewcommand{\thefigure}{S\arabic{figure}}
 %redefine the command that creates the equation no.
\setcounter{figure}{0}  % reset counter

\renewcommand{\thetable}{S\arabic{table}}
 %redefine the command that creates the equation no.
\setcounter{table}{0}

\section{Comparison of  metrological gains over the quantum standard limit with other experiments}

Our experiments generate multiparticle entangled states of up to 19 superconducting qubits through
the short-time nonlinear evolution of the system with the Hamiltonian in Eq.~({\color{magenta}1}) of the main text.
%The experimental device and its effective Hamiltonian are discussed in details in Section~\ref{sec:2}.
To characterise the useful entangled states for quantum metrology with superconducting qubits,
we measure the linear and nonlinear spin squeezing parameters (discussed in Section~\ref{sec:4})
and extract the Fisher information from the squared Hellinger distance (discussed in
Section~\ref{sec:5}), which all indicate the metrological gain over the quantum standard limit (SQL)
of the phase sensitivity $\Delta \theta_{\textrm{SQL}} \sim 1/\sqrt{N}$, with $N$ being the number of particles.

\begin{figure}[t]
\centering
\includegraphics[width=0.98\linewidth]{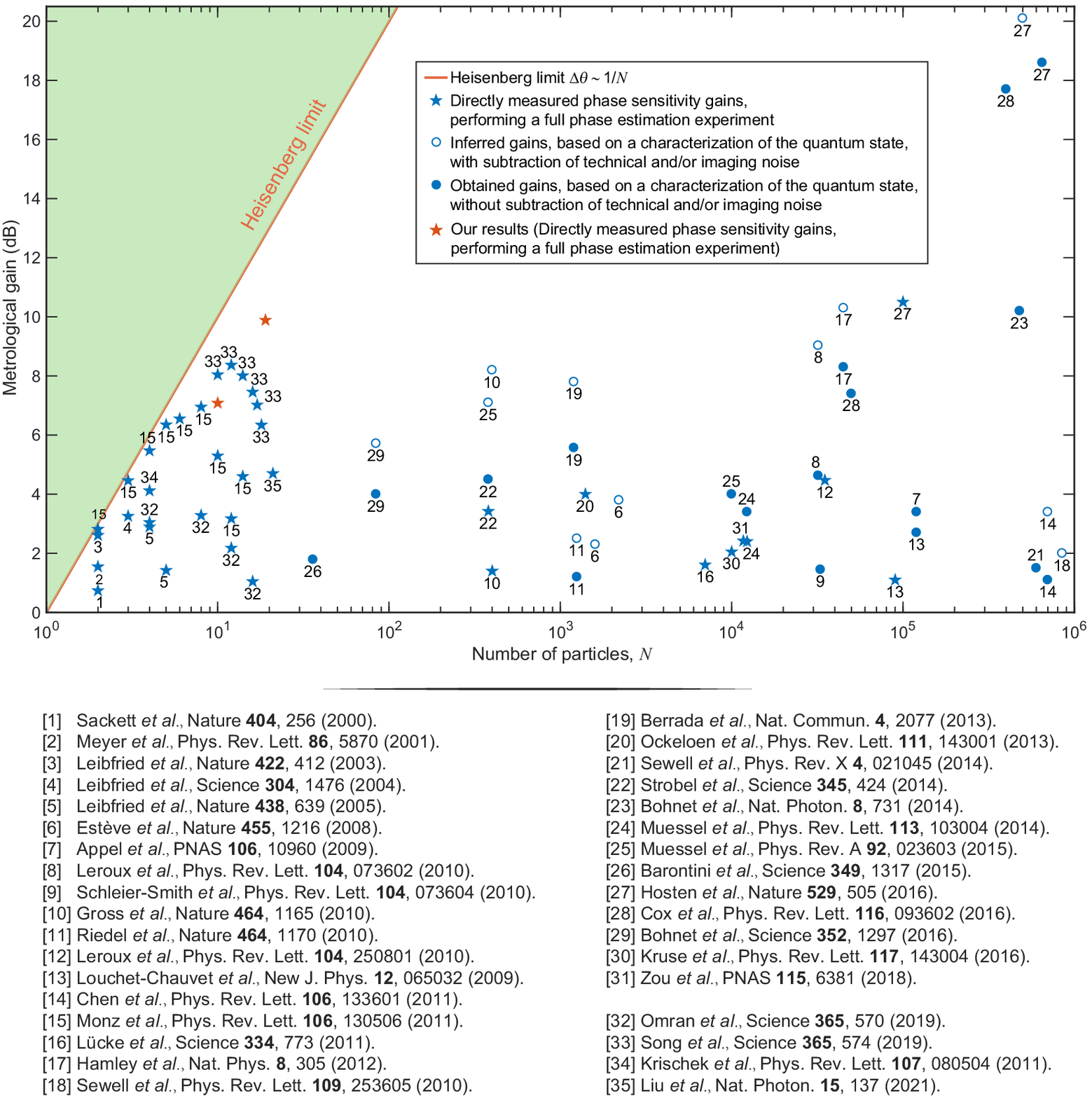}
\caption{\textbf{Metrological gains of the phase sensitivity over the standard quantum limit.}
Comparing the metrological gains of phase sensitivity $\Delta \theta$ over the standard quantum limit $\Delta\theta_\textrm{SQL}\sim1/\sqrt{N}$ with
other experiments, which are shown on logarithmic scales, $10\log_{10}(\Delta\theta_\textrm{SQL}/\Delta\theta)^2$~dB. The solid red line shows the Heisenberg limit $\Delta\theta_\textrm{HL}\sim1/{N}$.
Each symbol is accompanied by a number, corresponding to the reference list below, where
Refs.~\cite{Sackett2000,Meyer2001,Leibfried2003,Leibfried2004,Leibfried2005,Esteve:2008aa,Appel2009,
Leroux2010,Schleier-Smith2010,Gross:2010aa,Riedel:2010aa,Leroux2010a,LouchetChauvet2010,Chen2011,
Monz2011,Luecke2011,Hamley2012,Sewell2012,Berrada2013,Ockeloen2013,Sewell2014,Strobel2014,
Bohnet2014,Muessel2014,Muessel2015,Barontini2015,Hosten2016,Cox2016,Bohnet2016,Kruse2016,Zou2018}
are in the same order and displayed using the same data and symbols as in Ref.~\cite{Pezze2018}.
}\label{fs8}
\end{figure}

In our experiments, the Fisher information of non-Gaussian entangled states in the over-squeezed
regime reveals the largest metrological gains. In Fig.~\ref{fs8}, our experimental results,
$F/N=7.10^{+0.26}_{-0.28}$~dB with $N=10$ qubits, and $F/N=9.89^{+0.28}_{-0.29}$~dB  with
$N=19$ qubits, are compared with other experimental results
\cite{Sackett2000,Meyer2001,Leibfried2003,Leibfried2004,Leibfried2005,Esteve:2008aa,Appel2009,
Leroux2010,Schleier-Smith2010,Gross:2010aa,Riedel:2010aa,Leroux2010a,LouchetChauvet2010,Chen2011,
Monz2011,Luecke2011,Hamley2012,Sewell2012,Berrada2013,Ockeloen2013,Sewell2014,Strobel2014,
Bohnet2014,Muessel2014,Muessel2015,Barontini2015,Hosten2016,Cox2016,Bohnet2016,Kruse2016,Zou2018,
Omran2019,Song2019,Krischek2011,Liu2021},
obtained on different experimental platforms including cold/thermal atoms, trapped ions,
Bose-Einstein condensates, photonic systems, Rydberg atoms, and superconducting qubits.
Here, the experimental results in refs.~\cite{Sackett2000,Meyer2001,Leibfried2003,Leibfried2004,Leibfried2005,Esteve:2008aa,Appel2009,
Leroux2010,Schleier-Smith2010,Gross:2010aa,Riedel:2010aa,Leroux2010a,LouchetChauvet2010,Chen2011,
Monz2011,Luecke2011,Hamley2012,Sewell2012,Berrada2013,Ockeloen2013,Sewell2014,Strobel2014,
Bohnet2014,Muessel2014,Muessel2015,Barontini2015,Hosten2016,Cox2016,Bohnet2016,Kruse2016,Zou2018}
have been reviewed in Fig.~2 of ref.~\cite{Pezze2018} (with the same list number). In addition, our comparison also includes several recent
experimental results in refs.~\cite{Omran2019,Song2019,Krischek2011,Liu2021}.

Our work shows that superconducting qubits, with high-fidelity controls and long decoherence times,
are able to efficiently perform quantum metrology tasks, since our \emph{metrological gain with 19
superconducting qubits is larger than the ones obtained on other platforms with up to 10,000 particles}.
In addition, the ability to perform  \emph{single-shot readout measurement on each qubit} on the superconducting
processor also makes the detection of \emph{nonlinear} squeezing parameters and other quantum-information tasks
possible. These advantages indicate the potential of an all-to-all connected superconducting circuit
architecture for exploring quantum many-body physics, and also for practical applications
in quantum metrology and quantum information processing.

\section{Measurement of linear and nonlinear spin squeezing parameters}\label{sec:4}
\subsection{Optimisation of metrological squeezing parameters}
To estimate the phase $\theta$, imprinted on a state with a density matrix $\rho_\theta$,
by measuring an observable $\hat{X}$,  the phase sensitivity (for the unbiased estimation) can be given by \cite{Pezze2018}
\begin{align}
\Delta^2\theta=\frac{1}{\nu}\frac{(\Delta_{\rho_\theta}\hat{X})^2}{(\partial_\theta\langle\hat{X}\rangle_{\rho_\theta})^2}
\equiv\frac{\xi^2[\rho_\theta,\hat{X}]}{\nu},
\end{align}
where $\nu$ is the number of trials of the measurement.
Here, the coefficient
\begin{align}
\xi^2[\rho_\theta,\hat{X}]\equiv\frac{(\Delta_{\rho_\theta}\hat{X})^2}{(\partial_\theta\langle\hat{X}\rangle_{\rho_\theta})^{2}}
\end{align}
is defined as the metrological squeezing parameter of $\rho$ with respect to the observable $\hat{X}$ \cite{Gessner2019}.
We consider the case that the phase is generated by a unitary process $\rho_\theta=e^{-i\hat{G}\theta}\rho e^{i\hat{G}\theta}$
 by a generator $\hat{G}$, and the squeezing parameter, in the limit $\theta\rightarrow 0$, can be obtained as
\begin{align}
\xi^2[\rho,\hat{X},\hat{G}]=\frac{(\Delta_{\rho}\hat{X})^2}{|\langle[\hat{X},\hat{G}]\rangle_\rho|^2},
\end{align}
where we have used the fact that $(\Delta_{\rho_\theta}\hat{X})^2\rightarrow(\Delta_{\rho}\hat{X})^2$, and
$\partial_\theta\langle\hat{X}\rangle_{\rho_\theta}\rightarrow-i\langle[\hat{X},\hat{G}]\rangle_\rho$.

We then introduce  a family of $D$ accessible operators
\begin{align}
\hat{\mathbf{S}}=(\hat{S}_1,\hat{S}_2,\hat{S}_3,\cdots,\hat{S}_D),
\end{align}
with which the observable can be expressed as
\begin{align}
\hat{X}=\hat{S}_{\hat{m}}=\hat{m}\cdot\hat{\mathbf{S}},
\end{align}
with the unit vector $\hat{m}\in \mathbb{R}^D$.
The generator is assumed to be a linear collective spin operator
\begin{align}
\hat{G}=\hat{J}_{\hat{n}}=\hat{n}\cdot\hat{\mathbf{J}}
\end{align}
in the direction $\hat{n}\in \mathbb{R}^3$ with the family of  linear collective spin operators
\begin{align}
\hat{\mathbf{J}}\equiv(\hat{J}_x,\hat{J}_y,\hat{J}_z).
\end{align}
Thus, the optimal metrological squeezing parameter for this family of operators can be written as
\begin{align}
\xi^2_{\textrm{opt}}[\rho,\hat{\mathbf{S}}]=\min_{\hat{X}\in\textrm{span}(\hat{\mathbf{S}})}\min_{\hat{G}\in\textrm{span}(\hat{\mathbf{J}})} \xi^2[\rho,\hat{X},\hat{G}]=\min_{\hat{m}\in \mathbb{R}^D}\min_{\hat{n}\in\mathbb{R}^3}\frac{N(\Delta _{\rho}\hat{S}_{\hat{m}})^2}{|\langle[\hat{S}_{\hat{m}},\hat{J}_{\hat{n}}]\rangle_{\rho}|^2}=\frac{N}{\lambda_{\textrm{max}}(\tilde{\mathbb{M}}[\rho,\hat{\mathbf{S}}])},\label{omsp}
\end{align}
where the last equality [Eq.~(3) in the main text] is proved in ref.~\cite{Gessner2019}, $N$ is the number of qubits, and $\lambda_{\textrm{max}}(\tilde{\mathbb{M}}[\rho,\hat{\mathbf{S}}])$ is the
largest eigenvalue of a $3\times 3$ matrix $\tilde{\mathbb{M}}[\rho,\hat{\mathbf{S}}]$. The matrix $\tilde{\mathbb{M}}[\rho,\hat{\mathbf{S}}]$
contains the first three rows and columns of a $D\times D$ matrix as
\begin{align}
\mathbb{M}[\rho,\hat{\mathbf{S}}]=\mathbb{C}^T[\rho,\hat{\mathbf{S}}]\mathbb{V}^{-1}[\rho,\hat{\mathbf{S}}]\mathbb{C}[\rho,\hat{\mathbf{S}}],
\end{align}
where $\mathbb{V}[\rho,\hat{\mathbf{S}}]$ is the covariance matrix
(symmetric, $\mathbb{V}^T=\mathbb{V}$)  with elements:
\begin{align}
\mathbb{V}_{ij}[\rho,\hat{\mathbf{S}}]=\textrm{Cov}_\rho(\hat{S}_i,\hat{S}_j)=\frac{\langle\{\hat{S}_i,\hat{S}_j\}\rangle_\rho}{2}-\langle\hat{S}_i\rangle_\rho\langle\hat{S}_j\rangle_\rho,
\end{align}
and $\mathbb{C}[\rho,\hat{\bm{H}}]$ is the real-valued skew-symmetric commutator matrix
(asymmetric, $\mathbb{C}^T=-\mathbb{C}$) with elements:
\begin{align}
\mathbb{C}_{ij}[\rho,\hat{\mathbf{S}}]=-i\langle[\hat{S}_i,\hat{S}_j]\rangle_\rho.
\end{align}

\subsection{Linear Ramsey squeezing parameter}
To optimise the linear Ramsey squeezing parameter using Eq.~(\ref{omsp}), we consider the accessible
operators as  spanned  by a family of collective spin operators
\begin{align}
\hat{\mathbf{S}}_{(1)}\equiv\hat{\mathbf{J}}=(\hat{J}_x,\hat{J}_y,\hat{J}_z)_{D=3}.\label{three}
\end{align}
Then, the optimal spin squeezed parameter can be calculated with matrices
\begin{align}
\mathbb{V}_{(1)}%[\rho,\hat{\bm{H}}]
&=\left(
\begin{array}{ccc}
(\Delta_\rho\hat{J}_x)^2&\color{red}{\textrm{cov}_\rho(\hat{J}_x,\hat{J}_y)}&\color{red}{\textrm{cov}_\rho(\hat{J}_x,\hat{J}_z)}\\
\vdots&(\Delta_\rho\hat{J}_y)^2&\color{red}{\textrm{cov}_\rho(\hat{J}_y,\hat{J}_z)}\\
\cdots&\cdots&(\Delta_\rho\hat{J}_z)^2
\end{array}
\right),\label{1V}
\end{align}
\begin{align}
\mathbb{C}_{(1)}%[\rho,\hat{\bm{H}}]
&=\left(
\begin{array}{ccc}
0&\langle \hat{J}_z\rangle_\rho&-\langle \hat{J}_y\rangle_\rho\\
-\langle \hat{J}_z\rangle_\rho&0&\langle \hat{J}_x\rangle_\rho\\
\langle \hat{J}_y\rangle_\rho&-\langle \hat{J}_x\rangle_\rho&0
\end{array}
\right),\label{1C}
\end{align}
where, e.g., the covariance $\textrm{cov}_\rho(\hat{J}_x,\hat{J}_y)$ can be measured by the
single-shot readout measurement of the observable operator $\hat{J}_{xy}\equiv(\hat{J}_x+\hat{J}_y)/\sqrt{2}$ with
\begin{align}
\color{red}{\textrm{cov}_\rho(\hat{J}_x,\hat{J}_y)}&=\langle \hat{J}_{xy}^2\rangle_\rho-\frac{\langle \hat{J}^2_{x}\rangle_\rho+\langle \hat{J}^2_{y}\rangle_\rho}{2}-\langle \hat{J}_{x}\rangle_\rho\langle \hat{J}_{y}\rangle_\rho,\\
\color{red}{\textrm{cov}_\rho(\hat{J}_x,\hat{J}_z)}&=\langle \hat{J}_{zx}^2\rangle_\rho-\frac{\langle \hat{J}^2_{x}\rangle_\rho+\langle \hat{J}^2_{z}\rangle_\rho}{2}-\langle \hat{J}_{x}\rangle_\rho\langle \hat{J}_{z}\rangle_\rho,\\
\color{red}{\textrm{cov}_\rho(\hat{J}_y,\hat{J}_z)}&=\langle \hat{J}_{yz}^2\rangle_\rho-\frac{\langle \hat{J}^2_{y}\rangle_\rho+\langle \hat{J}^2_{z}\rangle_\rho}{2}-\langle \hat{J}_{y}\rangle_\rho\langle \hat{J}_{z}\rangle_\rho.
\end{align}

\begin{figure}[t]
\centering
\includegraphics[width=0.4\linewidth]{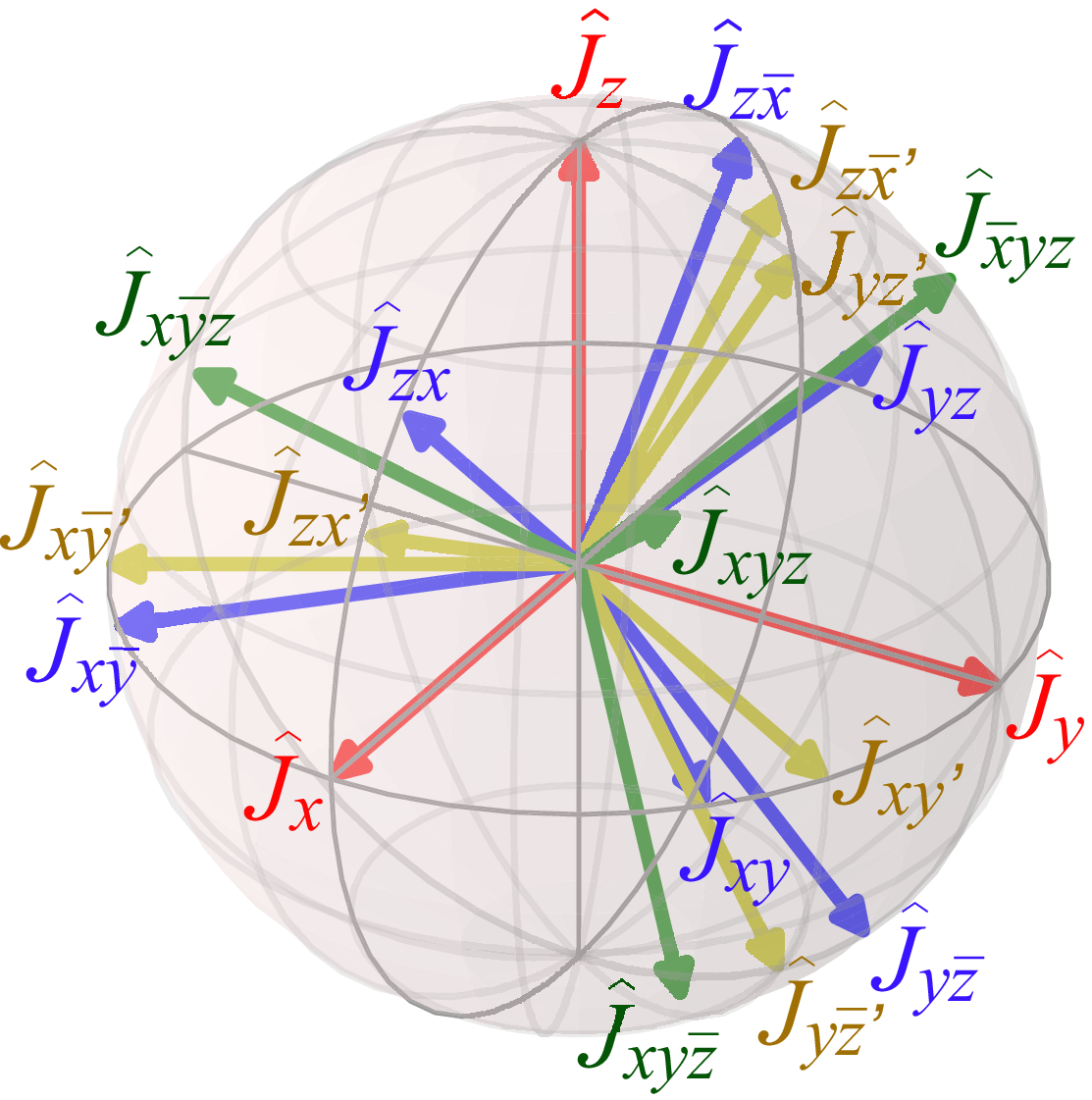}
\caption{\textbf{Directions of collective spin operators for single-shot readout measurements.}
The collective spin operators to obtain the
second-order nonlinear squeezing parameter via performing the single-shot readout measurement
on each superconducting qubit: $\{{\color{red}\hat{J}_x}, {\color{red}\hat{J}_y},
{\color{red}\hat{J}_z}, {\color{blue}\hat{J}_{xy}}, {\color{blue}\hat{J}_{yz}}, {\color{blue}\hat{J}_{zx}},
{\color{blue}\hat{J}_{x\bar{y}}}, {\color{blue}\hat{J}_{y\bar{z}}}, {\color{blue}\hat{J}_{z\bar{x}}},
{\color[rgb]{0.6,0.4,0.2}\hat{J}_{x{y}'}},
{\color[rgb]{0.6,0.4,0.2}\hat{J}_{y{z}'}},
{\color[rgb]{0.6,0.4,0.2}\hat{J}_{z{x}'}},
{\color[rgb]{0.6,0.4,0.2}\hat{J}_{x\bar{y}'}},
{\color[rgb]{0.6,0.4,0.2}\hat{J}_{y\bar{z}'}},
{\color[rgb]{0.6,0.4,0.2}\hat{J}_{z\bar{x}'}},
{\color[rgb]{0.0,0.5,0.0}\hat{J}_{xyz}},
{\color[rgb]{0.0,0.5,0.0}\hat{J}_{\bar{x}yz}},
{\color[rgb]{0.0,0.5,0.0}\hat{J}_{x\bar{y}z}},
{\color[rgb]{0.0,0.5,0.0}\hat{J}_{xy\bar{z}}}\}$,
as shown in Eqs.~(\ref{op1}--\ref{op2}). The unit vectors
for the directions of these collective spin operators are plotted in a unit sphere.
}\label{fs2}
\end{figure}

\subsection{Second-order nonlinear squeezing parameter}
For the second-order nonlinear squeezing parameter, we introduce a family of
$D=9$  linear and quadratic collective spin operators,
\begin{align}
\hat{\mathbf{S}}_{(2)}=(\hat{J}_x,\hat{J}_y,\hat{J}_z,\hat{J}_x^2,\hat{J}_y^2,\hat{J}_z^2,\hat{J}_{xy}^2,\hat{J}_{yz}^2,\hat{J}_{zx}^2)_{D=9},
\end{align}
with the single-shot readout measurements of the operators, as shown in Fig.~\ref{fs2} with direction vectors on a unit sphere.
\begin{align}
&{\color{red}\hat{J}_x},\hspace{0.2 in} {\color{red}\hat{J}_y},\hspace{0.2 in} {\color{red}\hat{J}_z}\label{op1}\\
&{\color{blue}\hat{J}_{xy}}=\frac{\hat{J}_x+\hat{J}_y}{\sqrt{2}},\hspace{0.2 in}  {\color{blue}\hat{J}_{yz}}=\frac{\hat{J}_y+\hat{J}_z}{\sqrt{2}},\hspace{0.2 in} {\color{blue}\hat{J}_{zx}}=\frac{\hat{J}_x+\hat{J}_z}{\sqrt{2}},\hspace{0.2 in}
{\color{blue}\hat{J}_{x\bar{y}}}=\frac{\hat{J}_x-\hat{J}_y}{\sqrt{2}},\hspace{0.2 in} {\color{blue}\hat{J}_{y\bar{z}}}=\frac{\hat{J}_y-\hat{J}_z}{\sqrt{2}},\hspace{0.2 in} {\color{blue}\hat{J}_{z\bar{x}}}=\frac{\hat{J}_z-\hat{J}_x}{\sqrt{2}},\\
&{\color[rgb]{0.6,0.4,0.2}\hat{J}_{x{y}'}}=\frac{\hat{J}_x+\sqrt{3}\hat{J}_y}{{2}},\hspace{0.2 in} {\color[rgb]{0.6,0.4,0.2}\hat{J}_{y{z}'}}=\frac{\hat{J}_y+\sqrt{3}\hat{J}_z}{{2}},\hspace{0.2 in} {\color[rgb]{0.6,0.4,0.2}\hat{J}_{z{x}'}}=\frac{\hat{J}_z+\sqrt{3}\hat{J}_x}{{2}},\\
&{\color[rgb]{0.6,0.4,0.2}\hat{J}_{x{\bar{y}'}}}=\frac{\hat{J}_x-\sqrt{3}\hat{J}_y}{{2}},\hspace{0.2 in} {\color[rgb]{0.6,0.4,0.2}\hat{J}_{y\bar{z}'}}=\frac{\hat{J}_y-\sqrt{3}\hat{J}_z}{{2}},\hspace{0.2 in} {\color[rgb]{0.6,0.4,0.2}\hat{J}_{{z}\bar{x}'}}=\frac{\hat{J}_z-\sqrt{3}\hat{J}_x}{{2}},\\
&{\color[rgb]{0.0,0.5,0.0}\hat{J}_{xyz}}=\frac{\hat{J}_x+\hat{J}_y+\hat{J}_z}{\sqrt{3}},\hspace{0.2 in} {\color[rgb]{0.0,0.5,0.0}\hat{J}_{\bar{x}{y}{z}}}=\frac{-\hat{J}_x+\hat{J}_y+\hat{J}_z}{\sqrt{3}},\hspace{0.2 in}
{\color[rgb]{0.0,0.5,0.0}\hat{J}_{{x}\bar{y}{z}}}=\frac{\hat{J}_x-\hat{J}_y+\hat{J}_z}{\sqrt{3}},\hspace{0.2 in} {\color[rgb]{0.0,0.5,0.0}\hat{J}_{{x}{y}\bar{z}}}=\frac{\hat{J}_x+\hat{J}_y-\hat{J}_z}{\sqrt{3}}.\label{op2}
\end{align}
The covariance matrix $\mathbb{V}_{(2)}$ for the second-order spin squeezing parameter is written as
\begin{align}
\mathbb{V}_{(2)}&=\left(
\begin{array}{c|ccc|ccc}
\color{white}{\vdots}& \textrm{cov}_\rho(J_x,J_x^2)& \color{red}{\textrm{cov}_\rho(\hat{J}_x,\hat{J}_y^2)}& \color{red}{\textrm{cov}_\rho(\hat{J}_x,\hat{J}_z^2)}&\color{blue}{ \textrm{cov}_\rho(\hat{J}_x,\hat{J}_{xy}^2)}&\color{blue}{ \textrm{cov}_\rho(\hat{J}_x,\hat{J}_{yz}^2)}&\color{blue}{ \textrm{cov}_\rho(\hat{J}_x,\hat{J}_{zx}^2)}\\
{\color{white}{\vdots}}\mathbb{V}_{(1)}{\color{white}{\vdots}}& \color{red}{\textrm{cov}_\rho(\hat{J}_y,\hat{J}_x^2)}& \textrm{cov}_\rho(\hat{J}_y,\hat{J}_y^2)& \color{red}{\textrm{cov}_\rho(J_y,J_z^2)}& \color{blue}{\textrm{cov}_\rho(J_y,J_{xy}^2)}& \color{blue}{\textrm{cov}_\rho(\hat{J}_y,\hat{J}_{yz}^2)}&\color{blue}{ \textrm{cov}_\rho(\hat{J}_y,\hat{J}_{zx}^2)}\\
\color{white}{\vdots}&\color{red}{ \textrm{cov}_\rho(\hat{J}_z,\hat{J}_x^2)}&\color{red}{ \textrm{cov}_\rho(\hat{J}_z,\hat{J}_y^2)}& \textrm{cov}_\rho(\hat{J}_z,\hat{J}_z^2)&\color{blue}{ \textrm{cov}_\rho(\hat{J}_z,\hat{J}_{xy}^2)}&\color{blue}{ \textrm{cov}_\rho(\hat{J}_z,\hat{J}_{yz}^2)}&\color{blue}{ \textrm{cov}_\rho(\hat{J}_z,\hat{J}_{zx}^2)}\\
\hline
\color{white}{\vdots}&(\Delta_\rho\hat{J}_x^2)^2&\color{red}{ \textrm{cov}_\rho(\hat{J}_x^2,\hat{J}_y^2)}&\color{red}{ \textrm{cov}_\rho(\hat{J}_x^2,\hat{J}_z^2)}&\color{blue}{ \textrm{cov}_\rho(\hat{J}_x^2,\hat{J}_{xy}^2)}&\color{blue}{ \textrm{cov}_\rho(\hat{J}_x^2,\hat{J}_{yz}^2)}& \color{blue}{\textrm{cov}_\rho(\hat{J}_x^2,\hat{J}_{zx}^2)}\\
\vdots&\vdots&(\Delta_\rho\hat{J}_y^2)^2& \color{red}{\textrm{cov}_\rho(\hat{J}_y^2,\hat{J}_z^2)}&\color{blue}{ \textrm{cov}_\rho(\hat{J}_y^2,\hat{J}_{xy}^2)}&\color{blue}{ \textrm{cov}_\rho(\hat{J}_y^2,\hat{J}_{yz}^2)}&\color{blue}{ \textrm{cov}_\rho(\hat{J}_y^2,\hat{J}_{zx}^2)}\\
\color{white}{\vdots}&\cdots&\cdots&(\Delta_\rho\hat{J}_z^2)^2& \color{blue}{\textrm{cov}_\rho(\hat{J}_z^2,\hat{J}_{xy}^2)}&\color{blue}{ \textrm{cov}_\rho(\hat{J}_z^2,\hat{J}_{yz}^2)}&\color{blue}{ \textrm{cov}_\rho(\hat{J}_z^2,\hat{J}_{zx}^2)}\\
\hline
\color{white}{\vdots}&&&&(\Delta_\rho\hat{J}_{xy}^2)^2&\color{magenta}{\textrm{cov}_\rho(\hat{J}_{xy}^2,\hat{J}_{yz}^2)}&\color{magenta}{\textrm{cov}_\rho(\hat{J}_{xy}^2,\hat{J}_{zx}^2)}\\
\vdots&\cdots&\cdots&\cdots&\vdots&(\Delta_\rho\hat{J}_{yz}^2)^2&\color{magenta}{\textrm{cov}_\rho(\hat{J}_{yz}^2,\hat{J}_{zx}^2)}\\
\color{white}{\vdots}&&&&\cdots&\cdots&(\Delta_\rho\hat{J}_{zx}^2)^2
\end{array}
\right),
\end{align}
with some of the elements being written in terms of the averages of measurable observables, as listed below: [We have shown the elements for the first three rows and
columns in Eqs.~(\ref{1V}).]
\begin{align}
 {\color{red}\textrm{cov}_\rho(\hat{J}_x^2,\hat{J}_y^2)}&=[2(\langle\hat{J}_{xy}^4\rangle_\rho+\langle\hat{J}_{x\bar{y}}^4\rangle_\rho)-\langle\hat{J}_x^4\rangle_\rho-\langle \hat{J}_y^4\rangle_\rho-3\langle\hat{J}_z^2\rangle_\rho+2\langle \hat{J}_y^2\rangle_\rho+2\langle\hat{J}_x^2\rangle_\rho]/6-\langle \hat{J}_x^2\rangle_\rho\langle\hat{J}_y^2\rangle_\rho,\\
  {\color{red}\textrm{cov}_\rho(\hat{J}_x^2,\hat{J}_z^2)}&=[2(\langle\hat{J}_{zx}^4\rangle_\rho+\langle\hat{J}_{z\bar{x}}^4\rangle_\rho)-\langle\hat{J}_x^4\rangle_\rho-\langle\hat{J}_z^4\rangle_\rho-3\langle\hat{J}_y^2\rangle_\rho+2\langle\hat{J}_x^2\rangle_\rho+2\langle\hat{J}_z^2\rangle_\rho]/6-\langle\hat{J}_x^2\rangle_\rho\langle\hat{J}_z^2\rangle_\rho,\\
   {\color{red}\textrm{cov}_\rho(\hat{J}_y^2,\hat{J}_z^2)}&=[2(\langle\hat{J}_{yz}^4\rangle_\rho+\langle\hat{J}_{y\bar{z}}^4\rangle_\rho)-\langle\hat{J}_y^4\rangle_\rho-\langle\hat{J}_z^4\rangle_\rho-3\langle\hat{J}_x^2\rangle_\rho+2\langle\hat{J}_y^2\rangle_\rho+2\langle\hat{J}_z^2\rangle_\rho]/6-\langle\hat{J}_y\rangle_\rho\langle\hat{J}_z^2\rangle_\rho,
\end{align}
and
\begin{align}
 {\color{red}\textrm{cov}_\rho(\hat{J}_x,\hat{J}_y^2)}&=[\sqrt{2}(\langle\hat{J}_{xy}^3\rangle_\rho+\langle\hat{J}_{x\bar{y}}^3\rangle_\rho)-\langle\hat{J}_x^3\rangle_\rho+\langle\hat{J}_x\rangle_\rho/2]/3-\langle\hat{J}_x\rangle_\rho\langle\hat{J}_y^2\rangle_\rho,\\
    {\color{red}\textrm{cov}_\rho(\hat{J}_y,\hat{J}_z^2)}&=[\sqrt{2}(\langle\hat{J}_{yz}^3\rangle_\rho+\langle\hat{J}_{y\bar{z}}^3\rangle_\rho)-\langle\hat{J}_y^3\rangle_\rho+\langle\hat{J}_y\rangle_\rho/2]/3-\langle\hat{J}_y\rangle_\rho\langle\hat{J}_z^2\rangle_\rho,\\
   {\color{red}\textrm{cov}_\rho(\hat{J}_z,\hat{J}_x^2)}&=[\sqrt{2}(\langle\hat{J}_{zx}^3\rangle_\rho+\langle\hat{J}_{z\bar{x}}^3\rangle_\rho)-\langle\hat{J}_z^3\rangle_\rho+\langle\hat{J}_z\rangle_\rho/2]/3-\langle\hat{J}_z\rangle_\rho\langle\hat{J}_x^2\rangle_\rho,\\
  {\color{red}\textrm{cov}_\rho(\hat{J}_x,\hat{J}_z^2)}&=[\sqrt{2}(\langle\hat{J}_{zx}^3\rangle_\rho-\langle\hat{J}_{z\bar{x}}^3\rangle_\rho)-\langle\hat{J}_x^3\rangle_\rho+\langle\hat{J}_x\rangle_\rho/2]/3-\langle\hat{J}_x\rangle_\rho\langle\hat{J}_z^2\rangle_\rho,\\
   {\color{red}\textrm{cov}_\rho(\hat{J}_y,\hat{J}_x^2)}&=[\sqrt{2}(\langle\hat{J}_{xy}^3\rangle_\rho-\langle\hat{J}_{x\bar{y}}^3\rangle_\rho)-\langle\hat{J}_y^3\rangle_\rho+\langle\hat{J}_y\rangle_\rho/2]/3-\langle\hat{J}_y\rangle_\rho\langle\hat{J}_x^2\rangle_\rho,\\
   {\color{red}\textrm{cov}_\rho(\hat{J}_z,\hat{J}_y^2)}&=[\sqrt{2}(\langle\hat{J}_{yz}^3\rangle_\rho-\langle\hat{J}_{y\bar{z}}^3\rangle_\rho)-\langle\hat{J}_z^3\rangle_\rho+\langle\hat{J}_z\rangle_\rho/2]/3-\langle\hat{J}_z\rangle_\rho\langle\hat{J}_y^2\rangle_\rho,
\end{align}
to further obtain that
\begin{align}
 {\color{blue}\textrm{cov}_\rho(\hat{J}_x,\hat{J}_{xy}^2)}&=\frac{\langle\hat{J}_x^3\rangle_\rho+{\color{red}\textrm{cov}_\rho(\hat{J}_x,\hat{J}_y^2)}}{2}+{\color{red}\textrm{cov}_\rho(\hat{J}_y,\hat{J}_x^2)}-\frac{\langle\hat{J}_y\rangle_\rho}{4}+\frac{\langle\hat{J}_x\rangle_\rho\langle\hat{J}_y^2\rangle_\rho}{2}+\langle\hat{J}_y\rangle_\rho\langle\hat{J}_x^2\rangle_\rho-\langle\hat{J}_x\rangle_\rho\langle\hat{J}_{xy}^2\rangle_\rho,\\
  {\color{blue}\textrm{cov}_\rho(\hat{J}_x,\hat{J}_{zx}^2)}&=\frac{\langle\hat{J}_x^3\rangle_\rho+{\color{red}\textrm{cov}_\rho(\hat{J}_x,\hat{J}_z^2)}}{2}+{\color{red}\textrm{cov}_\rho(\hat{J}_z,\hat{J}_x^2)}-\frac{\langle\hat{J}_z\rangle_\rho}{4}+\frac{\langle\hat{J}_x\rangle_\rho\langle_\rho\hat{J}_z^2\rangle_\rho}{2}+\langle\hat{J}_z\rangle_\rho\langle\hat{J}_x^2\rangle_\rho-\langle\hat{J}_x\rangle_\rho\langle\hat{J}_{zx}^2\rangle_\rho,\\
 {\color{blue}\textrm{cov}_\rho(\hat{J}_y,\hat{J}_{xy}^2)}&=\frac{\langle\hat{J}_y^3\rangle_\rho+{\color{red}\textrm{cov}_\rho(\hat{J}_y,\hat{J}_x^2)}}{2}+{\color{red}\textrm{cov}_\rho(\hat{J}_x,\hat{J}_y^2)}-\frac{\langle\hat{J}_x\rangle_\rho}{4}+\frac{\langle\hat{J}_y\rangle_\rho\langle\hat{J}_x^2\rangle_\rho}{2}+\langle\hat{J}_x\rangle_\rho\langle\hat{J}_y^2\rangle_\rho-\langle\hat{J}_y\rangle_\rho\langle\hat{J}_{xy}^2\rangle_\rho,\\
  {\color{blue}\textrm{cov}_\rho(\hat{J}_y,\hat{J}_{yz}^2)}&=\frac{\langle\hat{J}_y^3\rangle_\rho+{\color{red}\textrm{cov}_\rho(\hat{J}_y,\hat{J}_z^2)}}{2}+{\color{red}\textrm{cov}_\rho(\hat{J}_z,\hat{J}_y^2)}-\frac{\langle\hat{J}_z\rangle_\rho}{4}+\frac{\langle\hat{J}_y\rangle_\rho\langle\hat{J}_z^2\rangle_\rho}{2}+\langle\hat{J}_z\rangle_\rho\langle\hat{J}_y^2\rangle_\rho-\langle\hat{J}_y\rangle_\rho\langle\hat{J}_{yz}^2\rangle_\rho,\\
    {\color{blue}\textrm{cov}_\rho(\hat{J}_z,\hat{J}_{yz}^2)}&=\frac{\langle\hat{J}_z^3\rangle_\rho+{\color{red}\textrm{cov}_\rho(\hat{J}_z,\hat{J}_y^2)}}{2}+{\color{red}\textrm{cov}_\rho(\hat{J}_y,\hat{J}_z^2)}-\frac{\langle\hat{J}_y\rangle_\rho}{4}+\frac{\langle\hat{J}_z\rangle_\rho\langle\hat{J}_y^2\rangle_\rho}{2}+\langle\hat{J}_y\rangle_\rho\langle\hat{J}_z^2\rangle_\rho-\langle\hat{J}_z\rangle_\rho\langle\hat{J}_{yz}^2\rangle_\rho,\\
  {\color{blue}\textrm{cov}_\rho(\hat{J}_z,\hat{J}_{zx}^2)}&=\frac{\langle\hat{J}_z^3\rangle_\rho+{\color{red}\textrm{cov}_\rho(\hat{J}_z,\hat{J}_x^2)}}{2}+{\color{red}\textrm{cov}_\rho(\hat{J}_x,\hat{J}_z^2)}-\frac{\langle\hat{J}_x\rangle_\rho}{4}+\frac{\langle\hat{J}_z\rangle_\rho\langle\hat{J}_x^2\rangle_\rho}{2}+\langle\hat{J}_x\rangle_\rho\langle\hat{J}_z^2\rangle_\rho-\langle\hat{J}_z\rangle_\rho\langle\hat{J}_{zx}^2\rangle_\rho,
  \end{align}
  and
  \begin{align}
    {\color{blue}\textrm{cov}_\rho(\hat{J}_x,\hat{J}_{yz}^2)}=&\frac{{\color{red}\textrm{cov}_\rho(\hat{J}_x,\hat{J}_y^2+\hat{J}_z^2)}}{2}+\frac{3^{\frac{3}{2}}\langle_\rho \hat{J}_{xyz}^3\rangle_\rho-2^{\frac{3}{2}}(\langle \hat{J}_{xy}^3\rangle_\rho+\langle \hat{J}_{yz}^3\rangle_\rho+\langle \hat{J}_{zx}^3\rangle_\rho)+\langle \hat{J}_x^3\rangle_\rho+\langle \hat{J}_y^3\rangle_\rho+\langle \hat{J}_z^3\rangle_\rho}{6}\nonumber\\
    &+\frac{\langle \hat{J}_x\rangle_\rho(\langle \hat{J}_y^2\rangle_\rho+\langle \hat{J}_z^2\rangle_\rho)}{2}-\langle \hat{J}_x\rangle_\rho\langle \hat{J}_{yz}^2\rangle_\rho,\\
         {\color{blue}\textrm{cov}_\rho(\hat{J}_z,\hat{J}_{xy}^2)}=&\frac{{\color{red}\textrm{cov}_\rho(\hat{J}_z,\hat{J}_x^2+\hat{J}_y^2)}}{2}+\frac{3^{\frac{3}{2}}\langle \hat{J}_{xyz}^3\rangle_\rho-2^{\frac{3}{2}}(\langle \hat{J}_{xy}^3\rangle_\rho+\langle \hat{J}_{yz}^3\rangle_\rho+\langle \hat{J}_{zx}^3\rangle_\rho)+\langle \hat{J}_x^3\rangle_\rho+\langle \hat{J}_y^3\rangle_\rho+\langle \hat{J}_z^3\rangle_\rho}{6}\nonumber\\
    &+\frac{\langle \hat{J}_z\rangle(\langle \hat{J}_x^2\rangle_\rho+\langle \hat{J}_y^2\rangle_\rho)}{2}-\langle \hat{J}_z\rangle_\rho\langle \hat{J}_{xy}^2\rangle_\rho,\\
               {\color{blue}\textrm{cov}_\rho(\hat{J}_y,\hat{J}_{zx}^2)}=&\frac{{\color{red}\textrm{cov}_\rho(\hat{J}_y,\hat{J}_z^2+\hat{J}_x^2)}}{2}+\frac{3^{\frac{3}{2}}\langle \hat{J}_{xyz}^3\rangle_\rho-2^{\frac{3}{2}}(\langle \hat{J}_{xy}^3\rangle_\rho+\langle \hat{J}_{yz}^3\rangle_\rho+\langle \hat{J}_{zx}^3\rangle_\rho)+\langle \hat{J}_x^3\rangle_\rho+\langle \hat{J}_y^3\rangle_\rho+\langle \hat{J}_z^3\rangle_\rho}{6}\nonumber\\
    &+\frac{\langle \hat{J}_y\rangle_\rho(\langle \hat{J}_x^2\rangle_\rho+\langle \hat{J}_z^2\rangle_\rho)}{2}-\langle \hat{J}_y\rangle_\rho\langle \hat{J}_{zx}^2\rangle_\rho,
\end{align}
and
\begin{align}
 {\color{blue}\textrm{cov}_\rho(\hat{J}_x^2,\hat{J}_{xy}^2)}&=\frac{\langle \hat{J}_x^4\rangle_\rho+{\color{red}\textrm{cov}_\rho(\hat{J}_x^2,\hat{J}_y^2)}}{2}+\frac{3(\langle \hat{J}_{xy}^4\rangle_\rho-\langle \hat{J}_{x\bar{y}}^4\rangle_\rho)}{4}-\frac{\langle \hat{J}_{xy'}^4\rangle_\rho-\langle \hat{J}_{x\bar{y}'}^4\rangle_\rho}{\sqrt{3}}+\frac{\langle\hat{J}_x^2\rangle_\rho\langle\hat{J}_y^2\rangle_\rho}{2}-\langle\hat{J}_x^2\rangle_\rho\langle \hat{J}_{xy}^2\rangle_\rho,\\
  {\color{blue}\textrm{cov}_\rho(\hat{J}_y^2,\hat{J}_{yz}^2)}&=\frac{\langle\hat{J}_y^4\rangle_\rho+{\color{red}\textrm{cov}_\rho(\hat{J}_y^2,\hat{J}_z^2)}}{2}+\frac{3(\langle\hat{J}_{yz}^4\rangle_\rho-\langle\hat{J}_{y\bar{z}}^4\rangle_\rho)}{4}-\frac{\langle\hat{J}_{yz'}^4\rangle_\rho-\langle \hat{J}_{y\bar{z}'}^4\rangle_\rho}{\sqrt{3}}+\frac{\langle\hat{J}_y^2\rangle_\rho\langle\hat{J}_z^2\rangle_\rho}{2}-\langle\hat{J}_y^2\rangle_\rho\langle\hat{J}_{yz}^2\rangle_\rho,\\
   {\color{blue}\textrm{cov}_\rho(\hat{J}_z^2,\hat{J}_{zx}^2)}&=\frac{\langle\hat{J}_z^4\rangle_\rho+{\color{red}\textrm{cov}_\rho(\hat{J}_z^2,\hat{J}_x^2)}}{2}+\frac{3(\langle\hat{J}_{zx}^4\rangle_\rho-\langle\hat{J}_{z\bar{x}}^4\rangle_\rho)}{4}-\frac{\langle\hat{J}_{zx'}^4\rangle_\rho-\langle\hat{J}_{z\bar{x}'}^4\rangle_\rho}{\sqrt{3}}+\frac{\langle\hat{J}_z^2\rangle_\rho\langle\hat{J}_x^2\rangle_\rho}{2}-\langle\hat{J}_z^2\rangle_\rho\langle\hat{J}_{zx}^2\rangle_\rho,\\
  {\color{blue}\textrm{cov}_\rho(\hat{J}_x^2,\hat{J}_{zx}^2)}&=\frac{\langle\hat{J}_x^4\rangle_\rho+{\color{red}\textrm{cov}_\rho(\hat{J}_x^2,\hat{J}_z^2)}}{2}-\frac{\langle\hat{J}_{zx}^4\rangle_\rho-\langle J_{z\bar{x}}^4\rangle_\rho}{4}+\frac{\langle\hat{J}_{zx'}^4\rangle_\rho-\langle\hat{J}_{z\bar{x}'}^4\rangle_\rho}{\sqrt{3}}+\frac{\langle \hat{J}_z^2\rangle_\rho\langle \hat{J}_x^2\rangle_\rho}{2}-\langle \hat{J}_x^2\rangle_\rho\langle J_{zx}^2\rangle_\rho,\\
 {\color{blue}\textrm{cov}_\rho(\hat{J}_y^2,\hat{J}_{xy}^2)}&=\frac{\langle \hat{J}_y^4\rangle_\rho+{\color{red}\textrm{cov}_\rho(\hat{J}_x^2,\hat{J}_y^2)}}{2}-\frac{\langle \hat{J}_{xy}^4\rangle_\rho-\langle \hat{J}_{x\bar{y}}^4\rangle_\rho}{4}+\frac{\langle \hat{J}_{xy'}^4\rangle_\rho-\langle \hat{J}_{x\bar{y}'}^4\rangle_\rho}{\sqrt{3}}+\frac{\langle \hat{J}_x^2\rangle_\rho\langle \hat{J}_y^2\rangle_\rho}{2}-\langle \hat{J}_y^2\rangle_\rho\langle \hat{J}_{xy}^2\rangle_\rho,\\
 {\color{blue}\textrm{cov}_\rho(\hat{J}_z^2,\hat{J}_{yz}^2)}&=\frac{\langle \hat{J}_z^4\rangle_\rho+{\color{red}\textrm{cov}_\rho(\hat{J}_y^2,\hat{J}_z^2)}}{2}-\frac{\langle \hat{J}_{yz}^4\rangle_\rho-\langle \hat{J}_{y\bar{z}}^4\rangle_\rho}{4}+\frac{\langle \hat{J}_{yz'}^4\rangle_\rho-\langle \hat{J}_{y\bar{z}'}^4\rangle_\rho}{\sqrt{3}}+\frac{\langle \hat{J}_y^2\rangle_\rho\langle \hat{J}_z^2\rangle_\rho}{2}-\langle \hat{J}_z^2\rangle_\rho\langle \hat{J}_{yz}^2\rangle_\rho,
  \end{align}
  and
  \begin{align}
    {\color{blue}\textrm{cov}_\rho(\hat{J}_x^2,\hat{J}_{yz}^2)}=&\frac{{\color{red}\textrm{cov}_\rho(\hat{J}_x^2,\hat{J}_y^2+\hat{J}_z^2)}}{2}+\frac{3(\langle \hat{J}^4_{\bar{x}{y}{z}}\rangle_{\rho}+\langle \hat{J}_{xyz}^4\rangle_{\rho})}{8}+\frac{\langle \hat{J}_x^4\rangle_{\rho}+\langle \hat{J}_y^4\rangle_{\rho}+\langle \hat{J}_z^4\rangle_{\rho}}{12}+\frac{5(2\langle \hat{J}_{yz}^2\rangle_{\rho}-\langle \hat{J}_y^2\rangle_{\rho}-\langle \hat{J}_z^2\rangle_{\rho})}{12}\nonumber\\
    &-\frac{2\langle \hat{J}_{yz}^4\rangle_{\rho}+\langle \hat{J}_{xy}^4\rangle_{\rho}+\langle \hat{J}_{x\bar{y}}^4\rangle_{\rho}+\langle \hat{J}_{zx}^4\rangle_{\rho}+\langle \hat{J}_{z\bar{x}}^4\rangle_{\rho}}{6}+\frac{\langle \hat{J}_x^2\rangle_{\rho}(\langle \hat{J}_y^2\rangle_{\rho}+\langle \hat{J}_z^2\rangle_{\rho})}{2}-\langle \hat{J}_x^2\rangle_{\rho}\langle \hat{J}_{yz}^2\rangle_{\rho},\\
      {\color{blue}\textrm{cov}_\rho(\hat{J}_y^2,\hat{J}_{zx}^2)}=&\frac{{\color{red}\textrm{cov}_\rho(\hat{J}_y^2,\hat{J}_z^2+\hat{J}_x^2)}}{2}
      +\frac{3(\langle \hat{J}^4_{x\bar{y}{z}}\rangle_{\rho}+\langle \hat{J}_{xyz}^4\rangle_{\rho})}{8}
      +\frac{\langle \hat{J}_x^4\rangle_{\rho}+\langle \hat{J}_y^4\rangle_{\rho}+\langle \hat{J}_z^4\rangle_{\rho}}{12}+\frac{5(2\langle \hat{J}_{zx}^2\rangle_{\rho}-\langle \hat{J}_z^2\rangle_{\rho}-\langle \hat{J}_x^2\rangle_{\rho})}{12}\nonumber\\
    &
      -\frac{2\langle \hat{J}_{zx}^4\rangle_{\rho}+\langle \hat{J}_{xy}^4\rangle_{\rho}+\langle \hat{J}_{x\bar{y}}^4\rangle_{\rho}+\langle \hat{J}_{yz}^4\rangle_{\rho}+\langle \hat{J}_{y\bar{z}}^4\rangle_{\rho}}{6}+\frac{\langle \hat{J}_y^2\rangle_{\rho}(\langle \hat{J}_z^2\rangle_{\rho}+\langle \hat{J}_x^2\rangle_{\rho})}{2}-\langle \hat{J}_y^2\rangle_{\rho}\langle \hat{J}_{zx}^2\rangle_{\rho},\\
         {\color{blue}\textrm{cov}_\rho(\hat{J}_z^2,\hat{J}_{xy}^2)}=&\frac{{\color{red}\textrm{cov}_\rho(\hat{J}_z^2,\hat{J}_x^2+\hat{J}_y^2)}}{2}+\frac{3(\langle \hat{J}^4_{x{y}\bar{z}}\rangle_{\rho}+\langle \hat{J}_{xyz}^4\rangle_{\rho})}{8}
         +\frac{\langle \hat{J}_x^4\rangle_{\rho}+\langle \hat{J}_y^4\rangle_{\rho}+\langle \hat{J}_z^4\rangle_{\rho}}{12}
         +\frac{5(2\langle \hat{J}_{xy}^2\rangle_{\rho}-\langle \hat{J}_x^2\rangle_{\rho}-\langle \hat{J}_y^2\rangle_{\rho})}{12}\nonumber\\
    &-\frac{2\langle \hat{J}_{xy}^4\rangle_{\rho}+\langle \hat{J}_{yz}^4\rangle_{\rho}+\langle \hat{J}_{y\bar{z}}^4\rangle_{\rho}+\langle \hat{J}_{zx}^4\rangle_{\rho}+\langle \hat{J}_{z\bar{x}}^4\rangle_{\rho}}{6}+\frac{\langle \hat{J}_z^2\rangle_{\rho}(\langle \hat{J}_x^2\rangle_{\rho}+\langle \hat{J}_y^2\rangle_{\rho})}{2}-\langle \hat{J}_z^2\rangle_{\rho}\langle \hat{J}_{xy}^2\rangle_{\rho},
\end{align}
then to further obtain that
\begin{align}
 {\color{magenta}\textrm{cov}_\rho(\hat{J}_{xy}^2,\hat{J}_{yz}^2)}
    =&\frac{{\color{blue}\textrm{cov}_\rho(\hat{J}_x^2+\hat{J}_y^2,\hat{J}_{yz}^2)}+{\color{blue}\textrm{cov}_\rho(\hat{J}_y^2+\hat{J}_z^2,\hat{J}_{xy}^2)}}{2}-\frac{{\color{red}\textrm{cov}_\rho(\hat{J}_x^2,\hat{J}_y^2)}+{\color{red}\textrm{cov}_\rho(\hat{J}_y^2,\hat{J}_z^2)}+{\color{red}\textrm{cov}_\rho(\hat{J}_z^2,\hat{J}_x^2)}}{4}\nonumber\\
 &
 +\frac{3(\langle \hat{J}_{xyz}^4\rangle_{\rho}+\langle \hat{J}_{x\bar{y}z}^4\rangle_{\rho})}{8}+\frac{\langle \hat{J}_x^4\rangle_{\rho}-2\langle \hat{J}_y^4\rangle_{\rho}+\langle \hat{J}_z^4\rangle_{\rho}}{12}-\frac{2\langle \hat{J}_{zx}^4\rangle_{\rho}+\langle \hat{J}_{yz}^4\rangle_{\rho}+\langle \hat{J}_{y\bar{z}}^4\rangle_{\rho}+\langle \hat{J}_{xy}^4\rangle_{\rho}+\langle \hat{J}_{x\bar{y}}^4\rangle_{\rho}}{6}
 \nonumber\\
 &-\frac{5(2\langle \hat{J}_{zx}^2\rangle_{\rho}-\langle \hat{J}_x^2\rangle_{\rho}-\langle \hat{J}_z^2\rangle_{\rho})}{24}+\frac{(\langle \hat{J}_x^2\rangle_{\rho}+\langle \hat{J}_y^2\rangle_{\rho})\langle \hat{J}_{yz}^2\rangle_{\rho}+(\langle \hat{J}_y^2\rangle_{\rho}+\langle \hat{J}_z^2\rangle_{\rho})\langle \hat{J}_{xy}^2\rangle_{\rho}}{2}\nonumber\\
 &-\frac{\langle \hat{J}_x^2\rangle_{\rho}\langle \hat{J}_y^2\rangle_{\rho}+\langle \hat{J}_y^2\rangle_{\rho}\langle \hat{J}_z^2\rangle_{\rho}+\langle \hat{J}_z^2\rangle_{\rho}\langle \hat{J}_x^2\rangle_{\rho}}{4}-\langle \hat{J}_{xy}^2\rangle_{\rho}\langle \hat{J}_{yz}^2\rangle_{\rho},\\
  {\color{magenta}\textrm{cov}_\rho(\hat{J}_{xy}^2,\hat{J}_{zx}^2)}
    =&\frac{{\color{blue}\textrm{cov}_\rho(\hat{J}_x^2+\hat{J}_y^2,\hat{J}_{zx}^2)}+{\color{blue}\textrm{cov}_\rho(\hat{J}_z^2+\hat{J}_x^2,\hat{J}_{xy}^2)}}{2}-\frac{{\color{red}\textrm{cov}_\rho(\hat{J}_x^2,\hat{J}_y^2)}+{\color{red}\textrm{cov}_\rho(\hat{J}_y^2,\hat{J}_z^2)}+{\color{red}\textrm{cov}_\rho(\hat{J}_z^2,\hat{J}_x^2)}}{4}
\nonumber\\
 & +\frac{3(\langle \hat{J}_{xyz}^4\rangle_{\rho}+\langle \hat{J}_{\bar{x}{y}z}^4\rangle_{\rho})}{8}+\frac{-2\langle \hat{J}_x^4\rangle_{\rho}+\langle \hat{J}_y^4\rangle_{\rho}+\langle \hat{J}_z^4\rangle_{\rho}}{12}-\frac{2\langle \hat{J}_{yz}^4\rangle_{\rho}+\langle \hat{J}_{zx}^4\rangle_{\rho}+\langle \hat{J}_{z\bar{x}}^4\rangle_{\rho}+\langle \hat{J}_{xy}^4\rangle_{\rho}+\langle \hat{J}_{x\bar{y}}^4\rangle_{\rho}}{6}
 \nonumber\\
 &-\frac{5(2\langle \hat{J}_{yz}^2\rangle_{\rho}-\langle \hat{J}_y^2\rangle_{\rho}-\langle \hat{J}_z^2\rangle_{\rho})}{24}+\frac{(\langle \hat{J}_x^2\rangle_{\rho}+\langle \hat{J}_y^2\rangle_{\rho})\langle \hat{J}_{zx}^2\rangle_{\rho}+(\langle \hat{J}_z^2\rangle_{\rho}+\langle \hat{J}_x^2\rangle_{\rho})\langle \hat{J}_{xy}^2\rangle_{\rho}}{2}\nonumber\\
 &-\frac{\langle \hat{J}_x^2\rangle_{\rho}\langle \hat{J}_y^2\rangle_{\rho}+\langle \hat{J}_y^2\rangle_{\rho}\langle \hat{J}_z^2\rangle_{\rho}+\langle \hat{J}_z^2\rangle_{\rho}\langle \hat{J}_x^2\rangle_{\rho}}{4}-\langle \hat{J}_{xy}^2\rangle_{\rho}\langle \hat{J}_{zx}^2\rangle_{\rho},\\
 {\color{magenta}\textrm{cov}_\rho(\hat{J}_{yz}^2,\hat{J}_{zx}^2)}
    =&\frac{{\color{blue}\textrm{cov}_\rho(\hat{J}_y^2+\hat{J}_z^2,\hat{J}_{zx}^2)}+{\color{blue}\textrm{cov}_\rho(\hat{J}_z^2+\hat{J}_x^2,\hat{J}_{yz}^2)}}{2}-\frac{{\color{red}\textrm{cov}_\rho(\hat{J}_x^2,\hat{J}_y^2)}+{\color{red}\textrm{cov}_\rho(\hat{J}_y^2,\hat{J}_z^2)}+{\color{red}\textrm{cov}_\rho(\hat{J}_z^2,\hat{J}_x^2)}}{4}
 \nonumber\\
 &+\frac{3(\langle \hat{J}_{xyz}^4\rangle_{\rho}+\langle \hat{J}_{x{y}\bar{z}}^4\rangle_{\rho})}{8}+\frac{\langle \hat{J}_x^4\rangle_{\rho}+\langle \hat{J}_y^4\rangle_{\rho}-2\langle \hat{J}_z^4\rangle_{\rho}}{12}-\frac{2\langle \hat{J}_{xy}^4\rangle_{\rho}+\langle \hat{J}_{yz}^4\rangle_{\rho}+\langle \hat{J}_{y\bar{z}}^4\rangle_{\rho}+\langle \hat{J}_{zx}^4\rangle_{\rho}+\langle \hat{J}_{z\bar{x}}^4\rangle_{\rho}}{6}
 \nonumber\\
 &-\frac{5(2\langle \hat{J}_{xy}^2\rangle_{\rho}-\langle \hat{J}_x^2\rangle_{\rho}-\langle \hat{J}_y^2\rangle_{\rho})}{24}+\frac{(\langle \hat{J}_y^2\rangle_{\rho}+\langle \hat{J}_z^2\rangle_{\rho})\langle \hat{J}_{zx}^2\rangle_{\rho}+(\langle \hat{J}_z^2\rangle_{\rho}+\langle \hat{J}_x^2\rangle_{\rho})\langle \hat{J}_{yz}^2\rangle_{\rho}}{2}
 \nonumber\\
 &-\frac{\langle \hat{J}_x^2\rangle_{\rho}\langle \hat{J}_y^2\rangle_{\rho}+\langle \hat{J}_y^2\rangle_{\rho}\langle \hat{J}_z^2\rangle_{\rho}+\langle \hat{J}_z^2\rangle_{\rho}\langle \hat{J}_x^2\rangle_{\rho}}{4}-\langle \hat{J}_{yz}^2\rangle_{\rho}\langle \hat{J}_{zx}^2\rangle_{\rho}.
\end{align}
The  skew-symmetric commutator matrix is written as
\begin{align}
\mathbb{C}&=-i\langle\left(
\begin{array}{ccc|ccc|ccc}
0&[\hat{J}_x,\hat{J}_y]&[\hat{J}_x,\hat{J}_z]&0& \color{red}{[\hat{J}_x,\hat{J}_y^2]}& \color{red}{[\hat{J}_x,\hat{J}_z^2]}&\color{blue}{[ \hat{J}_x,\hat{J}_{xy}^2]}&\color{blue}{[\hat{J}_x,\hat{J}_{yz}^2]}&\color{blue}{[ \hat{J}_x,\hat{J}_{zx}^2]}\\
&0&[\hat{J}_y,\hat{J}_z]& \color{red}{[\hat{J}_y,\hat{J}_x^2]}& 0& \color{red}{[\hat{J}_y,\hat{J}_z^2]}& \color{blue}{[\hat{J}_y,\hat{J}_{xy}^2]}& \color{blue}{[\hat{J}_y,\hat{J}_{yz}^2]}&\color{blue}{ [\hat{J}_y,\hat{J}_{zx}^2]}\\
&&0&\color{red}{[ \hat{J}_z,\hat{J}_x^2]}&\color{red}{ [\hat{J}_z,\hat{J}_y^2]}&0&\color{blue}{ [\hat{J}_z,\hat{J}_{xy}^2]}&\color{blue}{[ \hat{J}_z,\hat{J}_{yz}^2]}&\color{blue}{[ \hat{J}_z,\hat{J}_{zx}^2]}\\
\hline
&&&0&\color{red}{ [\hat{J}_x^2,\hat{J}_y^2]}&\color{red}{ [\hat{J}_x^2,\hat{J}_z^2]}&\color{blue}{ [\hat{J}_x^2,\hat{J}_{xy}^2]}&\color{blue}{ [\hat{J}_x^2,\hat{J}_{yz}^2]}& \color{blue}{[\hat{J}_x^2,\hat{J}_{zx}^2]}\\
&&&&0& \color{red}{[\hat{J}_y^2,\hat{J}_z^2]}&\color{blue}{ [\hat{J}_y^2,\hat{J}_{xy}^2]}&\color{blue}{[ \hat{J}_y^2,\hat{J}_{yz}^2]}&\color{blue}{ [\hat{J}_y^2,\hat{J}_{zx}^2]}\\
&&&&&0& \color{blue}{[\hat{J}_z^2,\hat{J}_{xy}^2]}&\color{blue}{ [\hat{J}_z^2,\hat{J}_{yz}^2]}&\color{blue}{[ \hat{J}_z^2,\hat{J}_{zx}^2]}\\
\hline
&&&&&&0&\color{magenta}{[\hat{J}_{xy}^2,\hat{J}_{yz}^2]}&\color{magenta}{[\hat{J}_{xy}^2,\hat{J}_{zx}^2]}\\
&&&&&&&0&\color{magenta}{[\hat{J}_{yz}^2,\hat{J}_{zx}^2]}\\
&&&&&&&&0
\end{array}
\right)\rangle{}_{\rho},
\end{align}
with elements
\begin{align}
&-i\langle[\hat{J}_x,\hat{J}_y]\rangle_{\rho}=\langle \hat{J}_z\rangle_{\rho},\hspace{0.2 in}
-i\langle[\hat{J}_x,\hat{J}_z]\rangle_{\rho}=-\langle \hat{J}_y\rangle_{\rho},\hspace{0.2 in}
-i\langle[\hat{J}_y,\hat{J}_z]\rangle_{\rho}=\langle \hat{J}_x\rangle_{\rho},\\
&-i{\color{red}\langle[\hat{J}_x,\hat{J}_y^2]\rangle_{\rho}}=2\langle \hat{J}_{yz}^2\rangle_{\rho}-\langle \hat{J}_y^2\rangle_{\rho}-\langle \hat{J}_z^2\rangle_{\rho},\hspace{0.2 in} -i{\color{red}\langle[\hat{J}_x,\hat{J}_z^2]\rangle_{\rho}}=2\langle \hat{J}_{y\bar{z}}^2\rangle_{\rho}-\langle \hat{J}_y^2\rangle_{\rho}- \langle \hat{J}_z^2\rangle_{\rho},\\
& -i{\color{red}\langle[\hat{J}_y,\hat{J}_z^2]\rangle_{\rho}}=2\langle \hat{J}_{zx}^2\rangle_{\rho}-\langle \hat{J}_z^2\rangle_{\rho}-\langle \hat{J}_x^2\rangle_{\rho},\hspace{0.2 in} -i{\color{red}\langle[\hat{J}_y,\hat{J}_x^2]\rangle_{\rho}}=2\langle \hat{J}_{z\bar{x}}^2\rangle_{\rho}-\langle \hat{J}_z^2\rangle_{\rho}- \langle \hat{J}_x^2\rangle_{\rho},\\
&-i{\color{red}\langle[\hat{J}_z,\hat{J}_x^2]\rangle_{\rho}}=2\langle \hat{J}_{xy}^2\rangle_{\rho}-\langle \hat{J}_x^2\rangle_{\rho}-\langle  \hat{J}_y^2\rangle_{\rho}, \hspace{0.2 in}-i{\color{red}\langle[\hat{J}_z,\hat{J}_y^2]\rangle_{\rho}}=2\langle \hat{J}_{x\bar{y}}^2\rangle_{\rho}-\langle \hat{J}_x^2\rangle_{\rho}-\langle \hat{J}_y^2\rangle_{\rho},
\end{align}
and
\begin{align}
-i{\color{red}\langle[\hat{J}_x^2,\hat{J}_y^2]\rangle_{\rho}}&=2\sqrt{3}\langle \hat{J}_{xyz}^3\rangle_{\rho}-\frac{4\sqrt{2}(\langle \hat{J}_{xy}^3\rangle_{\rho}+\langle \hat{J}_{yz}^3\rangle_{\rho}+\langle \hat{J}_{zx}^3\rangle_{\rho})}{3}
+\frac{2(\langle \hat{J}_{x}^3\rangle_{\rho}+\langle \hat{J}_{y}^3\rangle_{\rho}+\langle \hat{J}_{z}^3\rangle_{\rho})}{3},\\
-i{\color{red}\langle[\hat{J}_x^2,\hat{J}_z^2]\rangle_{\rho}}&=i{\color{red}\langle[\hat{J}_x^2,\hat{J}_y^2]\rangle_{\rho}},\\
-i{\color{red}\langle[\hat{J}_y^2,\hat{J}_z^2]\rangle_{\rho}}&=-i{\color{red}\langle[\hat{J}_x^2,\hat{J}_y^2]\rangle_{\rho}},
\end{align}
to further obtain that
\begin{align}
&-i{\color{blue}\langle[\hat{J}_x,\hat{J}_{xy}^2]\rangle_{\rho}}=\frac{i}{2}{\color{red}\langle[\hat{J}_y,\hat{J}_{x}^2]\rangle_{\rho}}-\frac{i}{2}{\color{red}\langle[\hat{J}_x,\hat{J}_{y}^2]\rangle_{\rho}},
~~~~-i{\color{blue}\langle[\hat{J}_x,\hat{J}_{zx}^2]\rangle_{\rho}}=\frac{i}{2}{\color{red}\langle[\hat{J}_z,\hat{J}_{x}^2]\rangle_{\rho}}-\frac{i}{2}{\color{red}\langle[\hat{J}_x,\hat{J}_{z}^2]\rangle_{\rho}}, \\
&-i{\color{blue}\langle[\hat{J}_y,\hat{J}_{xy}^2]\rangle_{\rho}}=\frac{i}{2}{\color{red}\langle[\hat{J}_x,\hat{J}_{y}^2]\rangle_{\rho}}-\frac{i}{2}{\color{red}\langle[\hat{J}_y,\hat{J}_{x}^2]\rangle_{\rho}},
~~~~-i{\color{blue}\langle[\hat{J}_y,\hat{J}_{yz}^2]\rangle_{\rho}}=\frac{i}{2}{\color{red}\langle[\hat{J}_z,\hat{J}_{y}^2]\rangle_{\rho}}-\frac{i}{2}{\color{red}\langle[\hat{J}_y,\hat{J}_{z}^2]\rangle_{\rho}}, \\
&-i{\color{blue}\langle[\hat{J}_z,\hat{J}_{yz}^2]\rangle_{\rho}}=\frac{i}{2}{\color{red}\langle[\hat{J}_y,\hat{J}_{z}^2]\rangle_{\rho}}-\frac{i}{2}{\color{red}\langle[\hat{J}_z,\hat{J}_{y}^2]\rangle_{\rho}},
~~~~-i{\color{blue}\langle[\hat{J}_z,\hat{J}_{zx}^2]\rangle_{\rho}}=\frac{i}{2}{\color{red}\langle[\hat{J}_x,\hat{J}_{z}^2]\rangle_{\rho}}-\frac{i}{2}{\color{red}\langle[\hat{J}_z,\hat{J}_{x}^2]\rangle_{\rho}},
\end{align}
and
\begin{align}
&-i{\color{blue}\langle[\hat{J}_x,\hat{J}_{yz}^2]\rangle_{\rho}}=-\frac{i}{2}{\color{red}\langle[\hat{J}_x,\hat{J}_{y}^2]\rangle_{\rho}}-\frac{i}{2}{\color{red}\langle[\hat{J}_x,\hat{J}_{z}^2]\rangle_{\rho}}+\langle \hat{J}_z^2\rangle_{\rho}-\langle \hat{J}_y^2\rangle_{\rho},\\
&-i{\color{blue}\langle[\hat{J}_y,\hat{J}_{zx}^2]\rangle_{\rho}}=-\frac{i}{2}{\color{red}\langle[\hat{J}_y,\hat{J}_{z}^2]\rangle_{\rho}}-\frac{i}{2}{\color{red}\langle[\hat{J}_y,\hat{J}_{x}^2]\rangle_{\rho}}+\langle \hat{J}_x^2\rangle_{\rho}-\langle \hat{J}_z^2\rangle_{\rho},\\
&-i{\color{blue}\langle[\hat{J}_z,\hat{J}_{xy}^2]\rangle_{\rho}}=-\frac{i}{2}{\color{red}\langle[\hat{J}_z,\hat{J}_{x}^2]\rangle_{\rho}}-\frac{i}{2}{\color{red}\langle[\hat{J}_z,\hat{J}_{y}^2]\rangle_{\rho}}+\langle \hat{J}_y^2\rangle_{\rho}-\langle \hat{J}_x^2\rangle_{\rho},
\end{align}
and
\begin{align}
&-i{\color{blue}\langle[\hat{J}_x^2,\hat{J}_{xy}^2]\rangle_{\rho}}=\frac{2\sqrt{2}(\langle \hat{J}_{zx}^3\rangle_{\rho}+\langle \hat{J}_{z\bar{x}}^3\rangle_{\rho})}{3}-\frac{2\langle \hat{J}_z^3\rangle_{\rho}}{3}-\frac{\langle \hat{J}_z\rangle_{\rho}}{6}-\frac{i}{2}{\color{red}\langle[\hat{J}_x^2,\hat{J}_{y}^2]\rangle_{\rho}},\\
&-i{\color{blue}\langle[\hat{J}_y^2,\hat{J}_{yz}^2]\rangle_{\rho}}=\frac{2\sqrt{2}(\langle \hat{J}_{xy}^3\rangle_{\rho}+\langle \hat{J}_{x\bar{y}}^3\rangle_{\rho})}{3}-\frac{2\langle \hat{J}_x^3\rangle_{\rho}}{3}-\frac{\langle \hat{J}_x\rangle_{\rho}}{6}-\frac{i}{2}{\color{red}\langle[\hat{J}_y^2,\hat{J}_{z}^2]\rangle_{\rho}}, \\
&-i{\color{blue}\langle[\hat{J}_z^2,\hat{J}_{zx}^2]\rangle_{\rho}}=\frac{2\sqrt{2}(\langle \hat{J}_{yz}^3\rangle_{\rho}+\langle \hat{J}_{y\bar{z}}^3\rangle_{\rho})}{3}-\frac{2\langle \hat{J}_y^3\rangle_{\rho}}{3}-\frac{\langle \hat{J}_y\rangle_{\rho}}{6}-\frac{i}{2}{\color{red}\langle[\hat{J}_z^2,\hat{J}_{x}^2]\rangle_{\rho}},\\
&-i{\color{blue}\langle[\hat{J}_x^2,\hat{J}_{zx}^2]\rangle_{\rho}}=-\frac{2\sqrt{2}(\langle \hat{J}_{xy}^3\rangle_{\rho}-\langle \hat{J}_{x\bar{y}}^3\rangle_{\rho})}{3}+\frac{2\langle \hat{J}_y^3\rangle_{\rho}}{3}+\frac{\langle \hat{J}_y\rangle_{\rho}}{6}-\frac{i}{2}{\color{red}\langle[\hat{J}_x^2,\hat{J}_{z}^2]\rangle_{\rho}},\\
&-i{\color{blue}\langle[\hat{J}_y^2,\hat{J}_{xy}^2]\rangle_{\rho}}=-\frac{2\sqrt{2}(\langle \hat{J}_{yz}^3\rangle_{\rho}-\langle \hat{J}_{y\bar{z}}^3\rangle_{\rho})}{3}+\frac{2\langle \hat{J}_z^3\rangle_{\rho}}{3}+\frac{\langle \hat{J}_z\rangle_{\rho}}{6}-\frac{i}{2}{\color{red}\langle[\hat{J}_y^2,\hat{J}_{x}^2]\rangle_{\rho}},\\
&-i{\color{blue}\langle[\hat{J}_z^2,\hat{J}_{yz}^2]\rangle_{\rho}}=-\frac{2\sqrt{2}(\langle \hat{J}_{zx}^3\rangle_{\rho}-\langle \hat{J}_{z\bar{x}}^3\rangle_{\rho})}{3}+\frac{2\langle \hat{J}_x^3\rangle_{\rho}}{3}+\frac{\langle \hat{J}_x\rangle_{\rho}}{6}-\frac{i}{2}{\color{red}\langle[\hat{J}_z^2,\hat{J}_{y}^2]\rangle_{\rho}},
\end{align}
and
\begin{align}
&-i{\color{blue}\langle[\hat{J}_x^2,\hat{J}_{yz}^2]\rangle_{\rho}}=\frac{2\sqrt{2}(\langle \hat{J}_{zx}^3\rangle_{\rho}-\langle \hat{J}_{z\bar{x}}^3\rangle_{\rho}-\langle \hat{J}_{xy}^3\rangle_{\rho}-\langle \hat{J}_{x\bar{y}}^3\rangle_{\rho})}{3}-\frac{i}{2}{\color{red}\langle[\hat{J}_x^2,\hat{J}_{y}^2]\rangle_{\rho}}-\frac{i}{2}{\color{red}\langle[\hat{J}_x^2,\hat{J}_{z}^2]\rangle_{\rho}},\\
&-i{\color{blue}\langle[\hat{J}_y^2,\hat{J}_{zx}^2]\rangle_{\rho}}=\frac{2\sqrt{2}(\langle \hat{J}_{xy}^3\rangle_{\rho}-\langle \hat{J}_{x\bar{y}}^3\rangle_{\rho}-\langle \hat{J}_{yz}^3\rangle_{\rho}-\langle \hat{J}_{y\bar{z}}^3\rangle_{\rho})}{3}-\frac{i}{2}{\color{red}\langle[\hat{J}_y^2,\hat{J}_{x}^2]\rangle_{\rho}}-\frac{i}{2}{\color{red}\langle[\hat{J}_y^2,\hat{J}_{z}^2]\rangle_{\rho}},\\
&-i{\color{blue}\langle[\hat{J}_z^2,\hat{J}_{xy}^2]\rangle_{\rho}}=\frac{2\sqrt{2}(\langle \hat{J}_{yz}^3\rangle_{\rho}-\langle \hat{J}_{y\bar{z}}^3\rangle_{\rho}-\langle \hat{J}_{zx}^3\rangle_{\rho}-\langle \hat{J}_{z\bar{x}}^3\rangle_{\rho})}{3}-\frac{i}{2}{\color{red}\langle[\hat{J}_z^2,\hat{J}_{x}^2]\rangle_{\rho}}-\frac{i}{2}{\color{red}\langle[\hat{J}_z^2,\hat{J}_{y}^2]\rangle_{\rho}},
\end{align}
then to further obtain that
\begin{align}
-i{\color{magenta}\langle[\hat{J}_{xy}^2,\hat{J}_{yz}^2]\rangle_{\rho}}=
&\frac{-i{\color{blue}\langle[\hat{J}_x^2,\hat{J}_{yz}^2]\rangle_{\rho}}-i{\color{blue}\langle[\hat{J}_y^2,\hat{J}_{yz}^2]\rangle_{\rho}}+i{\color{blue}\langle[\hat{J}_y^2,\hat{J}_{xy}^2]\rangle_{\rho}}+i{\color{blue}\langle[\hat{J}_z^2,\hat{J}_{xy}^2]\rangle_{\rho}}}{2}%\\&
+\frac{i{\color{blue}\langle[\hat{J}_x^2,\hat{J}_{y}^2]\rangle_{\rho}}+i{\color{blue}\langle[\hat{J}_x^2,\hat{J}_{z}^2]\rangle_{\rho}}+i{\color{blue}\langle[\hat{J}_y^2,\hat{J}_{z}^2]\rangle_{\rho}}}{4}\nonumber
\\&+{\color{blue}\textrm{cov}_\rho(\hat{J}_y,\hat{J}_z^2+\hat{J}_x^2)}-\langle \hat{J}_y^3\rangle_{\rho}-\frac{\langle \hat{J}_y\rangle_{\rho}}{4}+\langle \hat{J}_y\rangle_{\rho}(\langle \hat{J}_z^2\rangle_{\rho}+\langle \hat{J}_x^2\rangle_{\rho}),\\
-i{\color{magenta}\langle[\hat{J}_{xy}^2,\hat{J}_{zx}^2]\rangle_{\rho}}=
&\frac{-i{\color{blue}\langle[\hat{J}_x^2,\hat{J}_{zx}^2]\rangle_{\rho}}-i\langle{\color{blue}[\hat{J}_y^2,\hat{J}_{zx}^2]\rangle_{\rho}+i\langle{\color{blue}[\hat{J}_z^2,\hat{J}_{xy}^2]}\rangle_{\rho}+i\langle{\color{blue}[\hat{J}_x^2,\hat{J}_{xy}^2]}\rangle_{\rho}}}{2}%\\&
+\frac{i{\color{blue}\langle[\hat{J}_x^2,\hat{J}_{z}^2]\rangle_{\rho}}+i{\color{blue}\langle[\hat{J}_y^2,\hat{J}_{z}^2]}\rangle_{\rho}+i{\color{blue}\langle[\hat{J}_y^2,\hat{J}_{x}^2]\rangle_{\rho}}}{4}\nonumber
\\&-{\color{blue}\textrm{cov}_\rho(\hat{J}_x,\hat{J}_y^2+\hat{J}_z^2)}+\langle \hat{J}_x^3\rangle_{\rho}+\frac{\langle \hat{J}_x\rangle_{\rho}}{4}-\langle \hat{J}_x\rangle_{\rho}(\langle \hat{J}_y^2\rangle_{\rho}+\langle \hat{J}_z^2\rangle_{\rho}),\\
-i{\color{magenta}\langle[\hat{J}_{yz}^2,\hat{J}_{zx}^2]\rangle_{\rho}}=
&\frac{-i{\color{blue}\langle[\hat{J}_y^2,\hat{J}_{zx}^2]\rangle_{\rho}}-i{\color{blue}\langle[\hat{J}_z^2,\hat{J}_{zx}^2]\rangle_{\rho}}+i{\color{blue}\langle[\hat{J}_x^2,\hat{J}_{yz}^2]\rangle_{\rho}}+i{\color{blue}\langle[\hat{J}_z^2,\hat{J}_{yz}^2]\rangle_{\rho}}}{2}%\\&
+\frac{i{\color{blue}\langle[\hat{J}_y^2,\hat{J}_{z}^2]\rangle_{\rho}}+i{\color{blue}\langle[\hat{J}_y^2,\hat{J}_{x}^2]\rangle_{\rho}}+i{\color{blue}\langle[\hat{J}_z^2,\hat{J}_{x}^2]\rangle_{\rho}}}{4}\nonumber
\\&+{\color{blue}\textrm{cov}_\rho(\hat{J}_z,\hat{J}_x^2+\hat{J}_y^2)}-\langle \hat{J}_z^3\rangle_{\rho}-\frac{\langle \hat{J}_z\rangle_{\rho}}{4}+\langle \hat{J}_z\rangle_{\rho}(\langle \hat{J}_x^2\rangle_{\rho}+\langle \hat{J}_y^2\rangle_{\rho}).
\end{align}

\subsection{Experimental details on measurement of squeezing parameters}
In Fig.~{\color{magenta}2}d of the main text, to calculate the linear spin squeezing parameter
$\xi_{\textrm{R}}^{2}$ at each time point $t$, we apply the experimental sequence shown in Fig.~{\color{red}1}d
of the main text, which is divided into four successive steps:
\begin{itemize}
\item[($i$)] The state preparation realised by Y$_{\frac{\pi}{2}}$ gates.
\item[($ii$)] The nonlinear evolution where all qubits are equally detuned.
\item[($iii$)] The rotation pulses to measure qubits at different directions.
\item[($iv$)] The final joint single-shot readout.
\end{itemize}
We performed experimental runs repetitively for about 200,000 times in total
for each linear collective spin operator, $\hat{J}_\beta$, listed in Eqs.~(\ref{op1}--\ref{op2}).
We then divided the results into 80 groups. For each group with $i=1,2,\cdots,80$
denoting the group index, we obtain the joint raw probabilities of 10 qubits
\begin{align}
	\mathcal{P}^{(i)}_\beta=\{P_{0...00},P_{0...01},P_{0...10},\cdots,P_{1...11}\},
\end{align}
and then perform the readout correction on them to obtain the corrected
probability, $\tilde{\mathcal{P}}^{(i)}_\beta$, after which the average of the observable, $\langle\hat{J}_\beta\rangle_{\rho_t}^{(i)}$,
can be calculated for each group. Following the
same process described above, we collect results for all the observables, $\{\hat{J}_\beta\}$, and calculate
the linear Ramsey squeezing parameter, $[\xi_{\textrm{R}}^{2}]^{(i)}$, using Eq.~({\color{magenta}3}) in the
main text. The mean value and error bar of the $\xi_{\textrm{R}}^{2}$ are estimated from these
80 groups of experimental data.

For the second-order nonlinear squeezing parameter,
$\xi_{\textrm{NL}}^{2}$, as it requires a much larger number of experimental repetitions to become stable,
which is time-consuming, we adopt a different method to estimate the error bar. From 84 groups
of experimental data in total, we randomly select 40 groups of them and average these selected
data (as a group labeled by $j$) to calculate the second-order nonlinear squeezing parameter,
$[\xi_{\textrm{NL}}^{2}]^{(j)}$. After repeating this process 10 times ($j=1,2,\cdots,10$),
we are able to estimate the error bar of the second-order nonlinear squeezing parameter by
calculating the standard deviation of $\{[\xi_{\textrm{NL}}^{2}]^{(1)},[\xi_{\textrm{NL}}^{2}]^{(2)},\cdots,[\xi_{\textrm{NL}}^{2}]^{(10)}\}$.

%{\color{red}Explain why we choose seven collective spin operators. Show experimental details and results}

%\begin{figure}[t]
%\centering
%\includegraphics[width=0.6\linewidth]{sfigure6.pdf}
%\caption{\textbf{Data of the $\mathbb{C}$ matrix.} \textbf{a--f},
%Matrix $\mathbb{C}[\rho_t,\hat{\mathbf{S}}_{\textrm{exp}}]$ experimentally measured at (\textbf{a}) $t=2$~ns, (\textbf{b}) $t=34$~ns, and (\textbf{c}) $t=50$~ns,
%compared with the numerical simulations (\textbf{d}), (\textbf{e}), and (\textbf{f}).
%}\label{fs3}
%\end{figure}
%
%\begin{figure}[t]
%\centering
%\includegraphics[width=0.6\linewidth]{sfigure7.pdf}
%\caption{\textbf{Data of the $\mathbb{V}$ matrix.} \textbf{a--f},
%Matrix $\mathbb{V}[\rho_t,\hat{\mathbf{S}}_{\textrm{exp}}]$ experimentally measured at (\textbf{a}) $t=2$~ns, (\textbf{b}) $t=34$~ns, and (\textbf{c}) $t=50$~ns,
%compared with the numerical simulations (\textbf{d}), (\textbf{e}), and (\textbf{f}).
%}\label{fs4}
%\end{figure}
%
%\begin{figure}[t]
%\centering
%\includegraphics[width=0.6\linewidth]{sfigure8.pdf}
%\caption{\textbf{Data of the $\mathbb{M}$ matrix.} \textbf{a--f},
%Matrix $\mathbb{M}[\rho_t,\hat{\mathbf{S}}_{\textrm{exp}}]$ experimentally measured at (\textbf{a}) $t=2$~ns, (\textbf{b}) $t=34$~ns, and (\textbf{c}) $t=50$~ns,
%compared with the numerical simulations (\textbf{d}), (\textbf{e}), and (\textbf{f}).
%}\label{fs5}
%\end{figure}

\section{Extraction of the Fisher information }\label{sec:5}
\subsection{Extraction of the Fisher information from the squared Hellinger distance}
Given the generator, $\hat{J}_{y}\equiv\sum_{j=1}^N\hat{\sigma}_j^y/2$, followed by an optimal angle,
$\alpha_{\textrm{opt}}$, of the rotation along the $x$-axis to maximise the Fisher information,
we imprint the phase $\theta$ on the state as
\begin{align}
\tilde{\rho}_t(\theta)=\exp(-i\hat{J}_{y}\theta)\exp(-i\hat{J}_{x}\alpha_{\textrm{opt}})\rho_t\exp(i\hat{J}_{x}\alpha_{\textrm{opt}})\exp(i\hat{J}_{y}\theta),
\end{align}
and measure each superconducting qubit by the single-shot readout measurement to obtain the probability distribution of the observable $\hat{J_z}\equiv\sum_{j=1}^N\hat{\sigma}_j^z/2$.
To extract the Fisher information \cite{Zhong2013}, we consider the squared Hellinger distance as
\begin{align}
d_{\textrm{H}}^2(\theta)=1-\mathcal{F}_{\textrm{C}}[\{P_z(0)\},\{P_z(\theta)\}]=1-\sum_z\sqrt{P_z(0)P_z(\theta)},
\end{align}
where the Bhattacharyya coefficient (classical fidelity) is written as
\begin{align}
\mathcal{F}_{\textrm{C}}[\{P_z(0)\},\{P_z(\theta)\}] =\sum_z\sqrt{P_z(0)P_z(\theta)},
\end{align}
with $P_z(\theta)$ being the probability distribution of the output $z=-\frac{N}{2},-\frac{N}{2}+1,\cdots,\frac{N}{2}-1, \frac{N}{2}$ of the observable
$\hat{J}_z$. For a small $\theta\rightarrow0$,
the  Taylor expansion of the squared Hellinger distance is given as \cite{Pezze2018}
\begin{align}
d_{\textrm{H}}^2(\theta)=\frac{F(0)}{8}\theta^2+\mathcal{O}(\theta^3),
\end{align}
where the Fisher information (divided by 8) can be regarded as the square of the speed of
the Hellinger distance
\begin{align}
\sqrt{F(0)/8}=v_{\textrm{H}}\equiv\left.\frac{\partial d_{\textrm{H}}(\theta)}{\partial \theta}\right|_{\theta=0}.
\end{align}
In theory, by maximising the squared Hellinger distance over all possible positive operator-valued measures (POVMs)
$\{\hat{E}\}$, the squared Bures distance can be obtained
\begin{align}
d_{\textrm{B}}^2(\theta)=\max_{\{\hat{E}\}}d_{\textrm{H}}^2(\theta)=1-\mathcal{F}_{\textrm{Q}}[\tilde{\rho}_t(0),\tilde{\rho}_t(\theta)],
\end{align}
where the Bures fidelity (quantum fidelity) between two states $\rho(0)$ and $\rho(\theta)$ reads
\begin{align}
\mathcal{F}_{\textrm{Q}}[\tilde{\rho}_t(0),\tilde{\rho}_t(\theta)]\equiv\textrm{Tr}[\sqrt{\sqrt{\tilde{\rho}_t(0)}\tilde{\rho}_t(\theta)\sqrt{\tilde{\rho}_t(0)}}].
\end{align}
The Taylor expansion of the squared Bures distance for $\theta\rightarrow0$ is given as \cite{BRAUNSTEIN1994}
\begin{align}
d_{\textrm{B}}^2(\theta)=\frac{F_{\textrm{Q}}[\tilde{\rho}_t(0)]}{8}\theta^2+\mathcal{O}(\theta^3)
\end{align}
where the quantum Fisher information (divided by 8) can be regarded as the square of the speed of the Bures distance
\begin{align}
\sqrt{F_{\textrm{Q}}[\tilde{\rho}_t(0)]/8}=v_{\textrm{B}}\equiv\left.\frac{\partial d_{\textrm{B}}(\theta)}{\partial\theta}\right|_{\theta=0},
\end{align}
and gives an achievable upper bound for the Fisher information for the optimal choice of the POVMs
\begin{align}
F_{\textrm{Q}}[\tilde{\rho}_t(0)]=\max_{\{\hat{E}\}}F(0).
\end{align}

\subsection{Experimental details for extracting the Fisher information}

In Fig.~{\color{magenta}4} of the main text, to obtain the Fisher information at time $t$, we apply the
experimental sequence in Fig.~{\color{magenta}3}a, which successively includes: ($i$) the state preparation pulse
Y$_{\frac{\pi}{2}}$, ($ii$) the nonlinear evolution $\exp({-i\hat{H}t}$), ($iii$) the optimisation rotation
X$_\alpha$, and ($iv$) the joint readouts in cases with and without the phase pulse Y$_\theta$ inserted
before the readouts. For each $\theta$ and $\alpha$, we obtain the joint readout probabilities of
19 qubits
\begin{align}
	\mathcal{P}(\theta,\alpha)=\{P_{0...00},P_{0...01},P_{0...10},\cdots,P_{1...11}\},
\end{align}
from which the probabilities  $\{P_{z}(\theta,\alpha)\}$ are extracted after performing
the readout correction on $\mathcal{P}(\theta,\alpha)$. The Fisher information
$F(\theta=0,\alpha)$ for different $\alpha$ can then be extracted from the squared
Hellinger distance of two states with and without the phase pulse Y$_\theta$ inserted before
the readouts using
\begin{align}
F(0,\alpha)\simeq\frac{8\times d_{\textrm{H}}^2(\theta,\alpha)}{\theta^2},
\end{align}
with $\theta$ being selected as a small value ($-0.05$~rad in our experiment). The quadratic curve
fitting of the square of the Hellinger distance versus the phase (Fig.~{\color{magenta}4}a in the main text)
fits the experimental data well for a relative small phase.

The optimised Fisher information is saturated by the optimal tomography angle
$\alpha_{\textrm{opt}}$ along the $x$-axis
\begin{align}
	F_{\textrm{opt}}(0)=\max_{\alpha}[F(0,{\alpha})].
\end{align}
To estimate the error bar, we perform about 600,000 experimental
runs and obtain about 240 groups of the probabilities $\{P_z(\theta,\alpha)\}^{(i)}$
for each $\alpha$ and $\theta$,
where $i$ denotes the group index. After performing the readout correction on these
probabilities, we randomly select 60 groups of them to calculate the Fisher information using the
method described above. We repeat this random sampling process  10 times to calculate
the final error bar of the Fisher information.
Note that  for $t=48$~ns, we only obtained 180 groups of probabilities from about 400,000
repetitive experimental runs in total, and we randomly selected 40 groups of them to calculate the error bar.

% In our experiments, after searching for the optimal angle $\alpha_{\textrm{opt}}$ of the rotation along the $x$-axis
% (Fig.~4B in the main text), we extract the Fisher information from the square of the Hellinger distance of two states using
% \begin{align}
% F_\alpha(0)\simeq\frac{8d_H^2(\theta)}{\theta^2},
% \end{align}
% when choosing a small phase $\theta=0.05$.

%
%\section{Numerical details}
%Numerical computations are performed using the \textsc{QuTiP} \cite{Johansson:2012aa,Johansson:2013aa}
%(the quantum toolbox in \textsc{Python}) and \textsc{NumPy}.
%The time evolutions of the system with a Hamiltonian [Eq.~({\color{magenta}1}) in the main text]  are numerically
%simulated using \textsc{QuTiP}’s
%master equation solver mesolve, where the parameters in Fig.~\ref{coupling matrix} are used.
%Because the evolution time is much shorter than the qubits’ energy relaxation time and dephasing time
%$t\ll T_1, T_2$, we neglect the effect of decoherence in simulations.
%
%

\bibliography{supp}